\documentclass[a4paper,11pt]{article}
\pdfoutput=1
\usepackage{jheppub}
\usepackage{epsfig}
\def\be{\begin{equation}}
\def\ee{\end{equation}}
\def\bea{\begin{array}}
\def\eea{\end{array}}
\def\beqa{\begin{eqnarray}}
\def\eeqa{\end{eqnarray}}
\def\beqas{\begin{eqnarray*}}
\def\eeqas{\end{eqnarray*}}

\def\bp{\begin{picture}}
\def\ep{\end{picture}}
\def\bc{\begin{center}}
\def\ec{\end{center}}
\def\bfig{\begin{figure}}
\def\efig{\end{figure}}

\def\bit{\begin{itemize}}
\def\eit{\end{itemize}}
\def\nn{\nonumber}
\def\f{\frac}

\def\[{\left[}
\def\]{\right]}
\def\({\left(}
\def\){\right)}

\def\..{\left.}
\def\.{\right.}
\def\tl{\tilde}
\def\ra{\rightarrow}
\def\la{\leftarrow}

\def\tm{\times}

\def\da{\dagger}

\def\la{\lambda}

\def\al{\alpha}

\def\ka{\kappa}

\def\ep{\epsilon}

\def\pa{\partial}
\def\pr{\prime}

\title{\boldmath Minimal Yukawa deflection of AMSB from the Kahler potential}
\author{Zhuang Li$^a$,}
\author{Fei Wang$^{a,1}$\note{Corresponding author.}}
\emailAdd{lizhuang@gs.zzu.edu.cn, feiwang@zzu.edu.cn}
\affiliation[a]{School of Physics, Zhengzhou University, 450001, ZhengZhou, P.R.China}
\abstract{ We propose a minimal Yukawa deflection scenario of AMSB from the Kahler potential through the Higgs-messenger mixing. Salient features of this scenario are discussed and realistic MSSM spectrum can be obtained. Such a scenario, which are very predictive, can solve the tachyonic slepton problem with less messenger species. Numerical results indicate that the LOSPs predicted by this scenario can not be good DM candidates. So it is desirable to extend this scenario with a Peccei-Quinn sector to solve the strong CP problem and at the same time provide new DM candidates. We propose a way to obtain a light axino mass in SUSY KSVZ axion model with Yukawa deflected anomaly mediation SUSY breaking mechanism. The axino can possibly be the LSP and act as a good DM candidate.
}

\begin{document}
\maketitle \indent
\newpage
\section{Introduction}

 Low energy supersymmetry(SUSY), which is one of the most attractive extensions of standard model(SM), can solve elegantly the gauge hierarchy problem by introducing various TeV scale superpartners. It can also realize successful gauge coupling unification  as well as providing proper dark matter (DM) candidates and baryogensis mechanisms. The Higgs scalar, which was discovered by the ATALS and CMS collaborations of LHC \cite{ATLAS:higgs,CMS:higgs} in 2012, lie miraculously in the small $'115-135'$ GeV window predicted by low energy SUSY. Despite of these impressive successes, low energy SUSY confronts many challenges from LHC experiments, especially the null search results of superpartners at LHC which constrain the gluino mass $m_{\tl{g}}$ to upon 2 TeV\cite{CMSSM1} and the top squark mass $m_{\tl{t}_1}$ to upon 1 TeV\cite{CMSSM2} in some simplified models. Such difficulties imply that the soft SUSY breaking parameters in low energy SUSY should have an intricate structure.

 It is well known that the low energy soft SUSY breaking parameters can be determined by the SUSY breaking mechanism in its UV completed theory. Therefore, it is important to survey which type of SUSY breaking mechanism can accommodate better the phenomenologically favored low energy soft SUSY breaking spectrum, for example, SUGRA\cite{SUGRA}, the gauge mediated SUSY breaking(GMSB)\cite{GMSB} mechanism or the anomaly mediated SUSY breaking(AMSB)\cite{AMSB} mechanism. The mSUGRA scenario, which is very predictive, was however disfavored by the global fit of the GAMBIT collaboration even if only the DM relic density upper bound is considered in addition to the muon $g-2$ anomaly\cite{GAMBIT}.  The discovered 125 GeV Higgs boson, which needs a large trilinear coupling $A_t$ for TeV scale stop masses, challenges ordinary GMSB scenarios with light stops in which the trilinear couplings are predict to vanish at the messenger scale\cite{GMSB:125}.

   Minimal AMSB, which contains only one free parameter $F_\phi\simeq m_{3/2}$, is insensitive to the UV theory\cite{AMSB:RGE} and predicts a flavor conservation soft SUSY breaking spectrum. Although it is very predictive, minimal AMSB predicts tachyonic slepton masses so that the minimal scenario must be extended\cite{tachyonslepton}. The most elegant solution from aesthetical point of view is the deflected AMSB\cite{AMSB:deflect,Nelson:2002sa}(dAMSB), in which
additional messengers are introduced to deflect the renormalization group equation (RGE) trajectory of AMSB and push the negative slepton squared masses to positive values \cite{dAMSB:example}.
On the other hand, $N\geq 4$ messenger species are always needed to generate positive slepton squared masses with a naturally negative deflection parameter,
possibly leading to strong gauge couplings below the GUT scale or Landau pole below the Planck scale. Besides, (radiative) natural SUSY spectrum\cite{rnaturalsusy} in general is not predicted by ordinary (d)AMSB scenarios. Additional gauge or Yukawa mediation contributions from messenger-matter interactions(mixing) in dAMSB can be advantageous in various aspects. Scenarios with such extensions had been studied in \cite{Fei:1508.01299,Fei:1602.01699,Fei:1703.10894,Fei:1704.05079,Fei:1710.06105} by one of the authors.

   Axion is the pseudo-Goldstone boson associated to the spontaneous breaking of the anomalous Peccei-Quinn(PQ) symmetry\cite{PQ} that is introduced to solve the $'strong~CP'$ problem of QCD. There are two types of popular $'invisible~axion'$ model in the literatures, the KSVZ model\cite{KSVZ} and the DFSZ model\cite{DFSZ}. KSVZ axion model, which can possibly appear in some SUSY breaking mechanisms with a messenger sector, introduces a PQ scalar and additional heavy quarks. Therefore, the induced topological term in its low energy effective theory is the only modification to the standard model Lagrangian. So KSVZ axion model, which predicts no unsuppressed tree-level couplings of axion to standard model matter fields, can evade some of the stringent experimental constraints and is well motivated theoretically. Axino is the fermionic SUSY partner of axion and can act as a cold DM candidate\cite{kawasaki}. Knowing the axino mass, on the other hand, is essential to determine whether the axino is the LSP or not. In the SUSY extension of KSVZ axion model, the axino mass is always of order $m_{3/2}$ in anomaly mediation scenarios\cite{axino:AMSB} and is heavier than ordinary MSSM sparticles. It is therefore interesting to see if the axino can possibly be the LSP and act as the DM particle in anomaly mediation scenarios.

In this paper, we propose to introduce minimal Yukawa deflection by the holomorphic terms in the Kahler potential. Predictive MSSM spectrum can be generated. We also find that the axino can be the LSP through proper Kahler deflection. This paper is organized as follows.
In Sec \ref{secII}, we propose our scenario and discuss the salient features of this scenario. In Sec \ref{secIII}, the soft SUSY parameters are given. The axino mass in an extension of our scenario with a PQ sector is discussed.  Our numerical results are given in Sec \ref{secIV}.  Sec \ref{secV} contains our conclusions.

\section{\label{secII} Minimal Yukawa Deflection From Kahler potential}

   Two approaches are proposed to deflect the AMSB trajectory with the presence of messengers, by pseudo-moduli field\cite{AMSB:deflect} or
holomorphic terms (for messengers) in the Kahler potential\cite{Nelson:2002sa}. Additional Yukawa deflection contributions from messenger-matter interactions(mixing) can also be introduced in both approaches\cite{Fei:1508.01299,Fei:1602.01699,Fei:1703.10894,Fei:1704.05079,Fei:1710.06105}. However, many salient features in scenario\cite{Fei:1710.06105} with the Yukawa deflection of the Kahler potential are obscured by the complicate structure of NMSSM. We show that Yukawa deflection from Kaher potential may take the minimal form through Higgs-messenger mixing and its salient features can be seen clearly in this scenario.

We introduce the following holomorphic terms involving the compensator field $\phi$ in the Kahler potential
\beqa
 K_h&\supseteq& \phi^\da\phi \[c_1 \bar{X}_{\bf \bar{5}}X_{\bf 5}+c_2 \bar{H}_{\bf \bar{5}} X_{\bf 5}+c_3 \bar{X}_{\bf \bar{5}}H_{\bf 5}
 +c_4 \bar{H}_{\bf \bar{5}} H_{\bf 5}+\sum\limits_{k=1}^{N_{S}} \ka_k \bar{S}_k S_k\]+h.c.~,
\eeqa
with $\bar{H}_{\bf \bar{5}}, H_{\bf 5}$ the Higgs superfields and $X_{\bf 5},\bar{X}_{\bf \bar{5}}$ the messenger superfields in ${\bf 5}$ and ${\bf \overline{5}}$ representations of SU(5), respectively. $\bar{S}_k,S_k$ are respectively the spectator messenger fields in ${\bf \overline{5}}$ and ${\bf 5}$ representations of SU(5), which are introduced to change only the gauge beta functions. Note that $\bar{S}_k,S_k$ cannot be the PQ messengers $Q_i,\tl{Q}_i$ introduced in KVSZ axion model because the PQ messenger combinations $\tl{Q}_iQ_i$ will carry non-trivial PQ charges and cannot appear as holomorphic terms in the Kahler potential.

  As any non-singular matrix can be diagonalized by bi-unitary transformations $M_{d}^\pr=U^\da M V$, the previous expressions can be rewritten in the matrix form
\beqa
 (\bar{X}_{\bf \bar{5}}~,\bar{H}_{\bf \bar{5}})&&\(\bea{cc}~c_1~&~c_2~\\~c_3~&~c_4~\eea\)\(\bea{c}X_{\bf 5}\\H_{\bf 5}\eea\)~,\nn\\
=(\bar{X}_{\bf \bar{5}}~,\bar{H}_{\bf \bar{5}})U^\da &&\(\bea{cc}~c_a~&~0~\\~0~&~c_b~\eea\) V\(\bea{c}X_{\bf 5}\\H_{\bf 5}\eea\)~,\nn\\
\equiv (\bar{X}^\pr_{\bf \bar{5}}~,\bar{H}^\pr_{\bf \bar{5}})&&\(\bea{cc}~c_a~&~0~\\~0~&~c_b~\eea\) \(\bea{c}X^\pr_{\bf 5}\\H^\pr_{\bf 5}\eea\)~,
\label{eigen:ab}
\eeqa
with the new mass eigenstates defined as
\beqa
    \(\bea{c}X^\pr_{\bf 5}\\H^\pr_{\bf 5}\eea\)\equiv V\(\bea{c}X_{\bf 5}\\H_{\bf 5}\eea\),~
~~~~\(\bea{c}\bar{X}^\pr_{\bf \bar{5}}\\\bar{H}^\pr_{\bf \bar{5}}\eea\)\equiv U^*\(\bea{c}\bar{X}_{\bf \bar{5}}\\ \bar{H}_{\bf \bar{5}}\eea\).
\label{matrix:higgs}
\eeqa
The eigenvalue of the Higgs fields corresponds to the (negligibly) smaller one. Requiring the MSSM Higgs fields $H^\pr,\bar{H}^\pr$ to stay light and keep naturalness, we require $c_a\gg c_b\approx 0$. So we can safely neglect the $c_b \bar{H}^\pr_{\bf \bar{5}} H^\pr_{\bf 5}$ term in the following discussions.  The coefficients need to satisfy the approximate relation
\beqa
c_1c_4\approx c_2c_3~.
\eeqa
 This requirement is trivially satisfied with $c_4=c_2=0$ or $c_4=c_3=0$. For example, with $c_4=c_3=0$, we can define
\beqa
\bar{X}_{\bf \bar{5}}^\pr&=&\f{1}{\sqrt{c_1^2+c_2^2}}\(~c_1\bar{X}_{\bf \bar{5}}+ c_2\bar{H}_{\bf \bar{5}} \),~~~~X^\pr_{\bf 5}=X_{\bf 5}~,\nn\\
\bar{H}_{\bf \bar{5}}^\pr&=&\f{1}{\sqrt{c_1^2+c_2^2}}\(-c_2\bar{X}_{\bf \bar{5}}+ c_1\bar{H}_{\bf \bar{5}}\),~~~~H^\pr_{\bf 5}=H_{\bf 5}~,
\eeqa
to rewrite the Kahler potential into
\beqa
K\supseteq   c_X \bar{X}_{\bf 5}^\pr X_{\bf 5} + h.c.~,~~~~{\rm with}~~~c_X\equiv \sqrt{c_1^2+c_2^2}.
\eeqa
 In this special case, the mixing angle between $\bar{X}_{\bf \bar{5}}$ and $\bar{H}_{\bf \bar{5}}$ are given by $\tan\theta={c_2}/{c_1}$.

 The holomorphic terms in the Kahler potential reduces to
\beqa
K\supseteq  \f{\phi^\da}{\phi} \[ c_a \bar{X}^\pr_{\bf \bar{5}}X^\pr_{\bf 5}\]+ h.c.~,~
\label{kahler:soft-X}
\eeqa
after the rescaling $\phi \Phi\ra \Phi$. With the F-term VEVs of the compensator fields $\phi=1+F_\phi\theta^2$, we have
\beqa
\label{kahlersoft}
{\cal L}\supseteq -c_a |F_\phi|^2  \bar{X}^\pr_{\bf \bar{5}} X^\pr_{\bf 5}+ F_\phi^\da\int d^2\theta c_a \bar{X}^\pr_{\bf \bar{5}} X^\pr_{\bf 5} +h.c.~~~.
\eeqa
 We thus arrive at the mass matrix for scalar fields $\bar{X}^{\pr}_{\bf \bar{5}},X^{\pr }_{\bf 5}$
 \beqa
 \label{tmatrix}
(~ \bar{X}^\pr_{\bf \bar{5}},~X^{\pr *}_{\bf 5}) \left(\bea{ccc}
 ~c_a^2& ~c_a\\ c_a& c_a^2
 \eea \right)\left(\bea{c}\bar{X}^{\pr *}_{\bf \bar{5}}\\X^{\pr }_{\bf 5}\eea\right)~.
\eeqa
We require $|c_a|>1$ so that the scalar components of messengers will not acquire lowest component VEVs.

The SUSY breaking effects can be taken into account by a spurion superfields $R$ with the resulting effective Lagrangian
\beqa
{\cal L}=\int d^2 \theta  c_a \bar{X}^\pr_{\bf \bar{5}} X^\pr_{\bf 5} R~,
\eeqa
and the spurion VEV as
\beqa
R\equiv M_R+\theta^2 F_R= F_\phi(1-\theta^2 F_\phi)~,
\eeqa
The deflection parameter is given by
\beqa
d\equiv\f{F_R}{M_R F_\phi}-1=-2.
\eeqa

 After integrating out the heavy messenger $\bar{X}^\pr_{\bf \bar{5}}, X^\pr_{\bf 5}$, we can obtain the low energy effective theory involving only the MSSM superfields. Besides, the heavy triplet parts within $\bar{H}^\pr_{\bf \bar{5}}, H^\pr_{\bf 5}$ are integrated out by assuming proper doublet-triplet splitting mechanism.

On the other hand, such spurion messenger-matter mixing can affect the AMSB RGE trajectory. The superpotential in terms of SU(5) representation
can be written as
\beqa
W&=& \tl{y}_{ab} \tl{P}_a \bar{H}_{\bf \bar{5}}Q_b+\tl{y}^\pr_{ab}Q_a Q_b H_{\bf 5} +R \[ c_a \bar{X}^\pr_{\bf \bar{5}}X^\pr_{\bf 5}\]~.
\eeqa
Here $\tl{P}_a$ and $Q_b$, with $a,b=1,2,3$ the family indices, are the standard model matter superfields in the $\bar{\bf 5}$ and ${\bf 10}$ representations of SU(5), respectively. At the messenger scale characterized by $F_\phi$, the superpotential will reduce to
\beqa
W&\supseteq &\tl{y}_{ab}^U  Q_{L,a} \tl{H}_u {U_{L,b}^c}-\tl{y}_{ab}^D  Q_{L,a} \tl{H}_d {D_{L,b}^c}-\tl{y}^E_{ab} {L_{L,a}} \tl{H}_d E_{L,b}^c~,\nn\\
&=&\tl{y}_{ab}^U \[(V^{-1})_{21}X_u +(V^{-1})_{22}H_u\] Q_{L,a} {U_{L,b}^c}~\nn\\
&-&\[\tl{y}_{ab}^D  Q_{L,a}{D_{L,b}^c}+\tl{y}^E_{ab} {L_{L,a}} E_{L,b}^c \]\[(U^{T})_{21}X_d +(U^{T})_{22}H_d\]~,
\label{tantheta}
\eeqa
which includes the couplings between the MSSM superfields and messengers.
Here $\tl{H}_u, \tl{H}_d$ correspond to the doublet components of $H_{\bf 5}$ and $\bar{H}_{\bf \bar{5}}$, respectively. The superfields $H_u, H_d$, on the other hand, correspond to the physical doublet components of $H^\pr_{\bf 5}$ and $\bar{H}^\pr_{\bf \bar{5}}$, respectively.

 We can rewrite the mixing matrix elements as
\beqa
(V^{-1})_{21}=\sin\theta_1,~(V^{-1})_{22}=\cos\theta_1~;~~~~(U^{T})_{21}=\sin\theta_2,~(U^{T})_{22}=\cos\theta_2~.
\eeqa
We should note that the Yukawa couplings $y^U_{ab}, {y}_{ab}^D,{y}_{ab}^E$ in the MSSM corresponds to
\beqa
y^U_{ab}=\tl{y}_{ab}^U\cos\theta_1~,~{y}_{ab}^D=\tl{y}_{ab}^D\cos\theta_2~,~{y}_{ab}^E=\tl{y}_{ab}^E\cos\theta_2~,
\eeqa
so we have the messenger-matter interaction strength
\beqa
\tl{y}_{ab}^U (V^{-1})_{21}=y^U_{ab}\tan\theta_1~,~~ \tl{y}_{ab}^D (U^{T})_{21}= {y}_{ab}^D\tan\theta_2~,~~
\tl{y}^E_{ab}(U^{T})_{21}={y}_{ab}^E\tan\theta_2~.
\eeqa
Appearance of scaled Yukawa couplings involving the tangent of the mixing parameters for messenger-matter interaction strengths is one of the salient features of this deflection scenario. They are required to be less than $\sqrt{4\pi}$ in the numerical studies.

  The effects of integrating out the messengers can be taken into account by
Giudice-Rattazi's wavefunction renormalization\cite{wavefunction} approach.
The messenger threshold $M_{mess}^2$ is replaced by spurious chiral superfields $X$ with $M_{mess}^2=X^\da X $.
 The soft gaugino masses at the messenger scale $F_\phi$ are given by
\beqa
M_{i}(M_{mess})&=& g_i^2\(\f{F_\phi}{2}\f{\pa}{\pa \ln\mu}-\f{d F_\phi}{2}\f{\pa}{\pa \ln |X|}\)\f{1}{g_i^2}(\mu,|X|)~,
\label{soft1}
\eeqa
with
\beqa
\f{\pa}{\pa \ln |X|} g_i(\al; |X|)=\f{\Delta b_i}{16\pi^2} g_i^3~.
\eeqa
 The trilinear soft terms can also be determined by the wavefunction renormalization approach because of the non-renormalization of the superpotential. After integrating out the messenger superfields, the wavefunction will depend on the messenger threshold. The trilinear soft terms at the messenger scale $F_\phi$ are given by
\beqa
A_0^{ijk}\equiv \f{A_{ijk}}{y_{ijk}}&=&\sum\limits_{i}\(-\f{F_\phi}{2}\f{\pa}{\pa\ln\mu}+\f{d F_\phi}{2}\f{\pa}{\pa\ln |X|}\) Z({\mu}; |X|)~,\nn\\
&=&\sum\limits_{i} \(-\f{F_\phi}{2} G_i^- +d F_\phi\f{\Delta G_i}{2}\)~,
\eeqa
with $\Delta G\equiv G^+-G^-$ the discontinuity across the messenger threshold. Here $'G^+(G^-)'$ denote respectively the anomalous dimension above (below) the messenger threshold. The soft scalar masses are given by
\beqa
\label{soft:scalar}
m^2_{soft}&=&-\left|-\f{F_\phi}{2}\f{\pa}{\pa\ln\mu}+\f{d F_\phi}{2}\f{\pa}{\pa\ln |X|}\right|^2 \ln \[Z_i(\mu,|X|)\]~,\\
&=&-\(\f{F_\phi^2}{4}\f{\pa^2}{\pa (\ln\mu)^2}+\f{d^2F^2_\phi}{4}\f{\pa}{\pa(\ln |X|)^2}
-\f{d F^2_\phi}{2}\f{\pa^2}{\pa\ln|X|\pa\ln\mu}\) \ln \[Z_i(\mu,|X|)\],\nn
\eeqa
at the messenger scale. Details of the expression involving the derivative of $\ln |X|$ can be found in \cite{chacko,shih,Fei:1508.01299,Fei}.

\section{\label{secIII}The soft SUSY breaking parameters}
We will discuss the consequence of Yukawa deflection from $H_u$( or $H_d$)-messenger mixing in the Kahler potential, respectively.
The soft SUSY breaking parameters at the scale $F_\phi$ after integrating out the messengers can be calculated with the
formulas from eqn.(\ref{soft1}) to eqn.(\ref{soft:scalar}).

\subsection{Scenario I: $H_u$-Messenger Mixing}
 This scenario corresponds to $\tan\theta_2=0$ in eqn.(\ref{tantheta}).
\bit
\item The gaugino masses are given as
\beqa
M_i=-F_\phi\f{\al_i(\mu)}{4\pi}\[b_i-(-2)\Delta b_i\]~,
\label{gaugino1}
\eeqa
with
\beqa
~(b_1~,b_2~,~b_3)&=&(\f{33}{5},~1,-3)~,
\eeqa
and the changes of $\beta$-function for the gauge couplings
\beqa
\Delta(b_1~,b_2~,~b_3)&=&(~1+N_{S},~1+N_{S},~1+N_S).
\eeqa
\item  The non-vanishing trilinear couplings are given as
\beqa
A_t&=&\f{F_\phi}{16\pi^2}\[\tl{G}_{y_t}-(-2)3y_t^2\tan^2\theta_1\]~,\nn\\
A_b&=&\f{F_\phi}{16\pi^2}\[\tl{G}_{y_b}-(-2)y_t^2\tan^2\theta_1\]~,\nn\\
A_\tau&=&\f{F_\phi}{16\pi^2}\tl{G}_{y_\tau}~,
\label{at1}\eeqa
with the beta function of the Yukawa couplings
\beqa
\tl{G}_{y_t}&=&6y_t^2+y_b^2-(\f{16}{3}g_3^2+3g_2^2+\f{13}{15}g_1^2)~,\nn\\
\tl{G}_{y_b}&=&y_t^2+6y_b^2+y_\tau^2-(\f{16}{3}g_3^2+3g_2^2+\f{7}{15}g_1^2)~,\nn\\
\tl{G}_{y_\tau}&=&3y_b^2+4y_\tau^2-(3g_2^2+\f{9}{5}g_1^2)~,
\eeqa
and the discontinuity of the anomalous dimensions
\beqa
\Delta \tl{G}_{Q_3}&=&y_t^2\tan^2\theta_1~,~~~~~~
\Delta \tl{G}_{t_L^c}=2 y_t^2\tan^2\theta_1~.
\eeqa

\item  The scalar soft parameters are given by
\beqa
m^2_{{H}_u}~~&=&\f{F_\phi^2}{16\pi^2}\[\f{3}{2}G_2\al^2_2+\f{3}{10}G_1\al^2_1\]
+\f{F_\phi^2}{(16\pi^2)^2}\[ 3y_t^2\tl{G}_{y_t}\]~,~\nn\\
m^2_{{H}_d}~~&=&\f{F_\phi^2}{16\pi^2}\[\f{3}{2}G_2\al^2_2+\f{3}{10}G_1\al^2_1\]
+\f{F_\phi^2}{(16\pi^2)^2}\[3y_b^2\tl{G}_{y_b}+y_\tau^2\tl{G}_{y_\tau}\]~,~\nn\\
m^2_{\tl{Q}_{L,a}}&=&\f{F_\phi^2}{16\pi^2}\[\f{8}{3} G_3 \al^2_3+\f{3}{2}G_2\al^2_2+\f{1}{30}G_1\al^2_1\]+\delta_{a,3}\f{F_\phi^2}{(16\pi^2)^2}\[y_t^2\tl{G}_{y_t}+y_b^2\tl{G}_{y_b}\]
+\delta_{a,3}\Delta m^2_{\tl{Q}_{L,3}}~,\nn\\
m^2_{\tl{U}^c_{L,a}}&=&\f{F_\phi^2}{16\pi^2}\[\f{8}{3} G_3 \al^2_3+\f{8}{15}G_1\al^2_1\]
+\delta_{a,3}\f{F_\phi^2}{(16\pi^2)^2}\[2y_t^2\tl{G}_{y_t}\]+\delta_{a,3}\Delta m^2_{\tl{U}^c_{L,3}} ~,\nn\\
m^2_{\tl{D}^c_{L;a}}&=&\f{F_\phi^2}{16\pi^2}\[\f{8}{3} G_3 \al^2_3+\f{2}{15}G_1\al^2_1\]
+\delta_{a,3}\f{F_\phi^2}{(16\pi^2)^2}\[2y_b^2\tl{G}_{y_b}\]~,~\nn\eeqa
\beqa
m^2_{\tl{L}_{L;a}}&=&\f{F_\phi^2}{16\pi^2}\[\f{3}{2}G_2\al_2^2+\f{3}{10}G_1\al_1^2\]
+\delta_{a,3}\f{F_\phi^2}{(16\pi^2)^2}\[y_\tau^2\tl{G}_{y_\tau}\]~,~\nn\\
m^2_{\tl{E}_{L;a}^c}&=&\f{F_\phi^2}{16\pi^2}\f{6}{5}G_1\al_1^2
+\delta_{a,3}\f{F_\phi^2}{(16\pi^2)^2}\[2y_\tau^2\tl{G}_{y_\tau}\]~,
\eeqa
with
\beqa
G_i&=&-b_i~,~~~~~~~~~~
(b_1,b_2,b_3)=(\f{33}{5},1,-3)~,
\eeqa
and Yukawa deflection contributions
\beqa
\Delta m^2_{\tl{Q}_{L,3}}&=&\f{d^2 F_\phi^2}{(16\pi^2)^2}\[y_{Q_3 X_u \tl{t}_R }^2G^+_{Q_3 X_u \tl{t}_R}\]=\f{d^2 F_\phi^2}{(16\pi^2)^2}\[y_t^2\tan^2\theta_1G^+_{Q_3 X_u \tl{t}_R}\]~,\nn\\
\Delta m^2_{\tl{U}^c_{L,3}} &=&\f{d^2 F_\phi^2}{(16\pi^2)^2}\[2y_{Q_3 X_u \tl{t}_R }^2G^+_{Q_3 X_u \tl{t}_R}\]=\f{d^2 F_\phi^2}{(16\pi^2)^2}\[2y_t^2\tan^2\theta_1G^+_{Q_3 X_u \tl{t}_R}\]~.
\eeqa
Here $d=-2$ and $\delta_{a,3}$ is the Kronecker delta. The beta function for $y_{Q_3 X_u \tl{t}_R}$ upon
the messenger threshold $F_\phi$ is given by
\beqa
G_{Q_3 X_u \tl{t}_R}^+&=&3y_t^2+y_b^2+6y_t^2\tan^2\theta_1-\f{16}{3}g_3^2-{3}g_2^2-\f{13}{15}g_1^2~.
\eeqa

%%\beqa G_{Q_3}^+&=&y_t^2(1+\tan^2\theta_1)+y_b^2-\f{8}{3}g_3^2-\f{3}{2}g_2^2-\f{1}{30}g_1^2~,\\
%% G_{t_R}^+&=&2y_t^2(1+\tan^2\theta_1)-\f{8}{3}g_3^2-\f{8}{15}g_1^2~,\\
%%% G_{X_u}^+&=&3y_t^2 \tan^2\theta_1-\f{3}{2}g_2^2-\f{3}{10}g_1^2~, \eeqa

\eit
\subsection{Scenario II: $H_d$-Messenger Mixing}
This scenario corresponds to $\tan\theta_1=0$ in eqn.(\ref{tantheta}). Similar to scenario I, the soft SUSY breaking parameters at the scale $F_\phi$ after integrating out the messengers can be readily calculated.
\bit
\item The gaugino masses are given as
\beqa
M_i=-F_\phi\f{\al_i(\mu)}{4\pi}\[b_i-(-2)\Delta b_i\]~,
\label{gaugino2}
\eeqa
with
\beqa
~(b_1~,b_2~,~b_3)&=&(\f{33}{5},~1,-3)~,
\eeqa
and the changes of $\beta$-function for the gauge couplings
\beqa
\Delta(b_1~,b_2~,~b_3)&=&(~1+N_S,~1+N_S,~1+N_S).
\eeqa
\item  The non-vanishing trilinear couplings are given as
\beqa
A_t&=&\f{F_\phi}{16\pi^2}\[\tl{G}_{y_t}-(-2)y_b^2\tan^2\theta_2\]~,\nn\\
A_b&=&\f{F_\phi}{16\pi^2}\[\tl{G}_{y_b}-(-2)3y_b^2\tan^2\theta_2\]~,\nn\\
A_\tau&=&\f{F_\phi}{16\pi^2}\[\tl{G}_{y_\tau}-(-2)3y_\tau^2\tan^2\theta_2\]~,
\label{at2}
\eeqa
with the beta function of the Yukawa couplings
\beqa
\tl{G}_{y_t}&=&6y_t^2+y_b^2-(\f{16}{3}g_3^2+3g_2^2+\f{13}{15}g_1^2)~,\nn\\
\tl{G}_{y_b}&=&y_t^2+6y_b^2+y_\tau^2-(\f{16}{3}g_3^2+3g_2^2+\f{7}{15}g_1^2)~,\nn\\
\tl{G}_{y_\tau}&=&3y_b^2+4y_\tau^2-(3g_2^2+\f{9}{5}g_1^2)~,
\eeqa
and the discontinuity of the anomalous dimension
\beqa
\Delta \tl{G}_{Q_3}&=&y_b^2\tan^2\theta_2~,~~~~~~
\Delta \tl{G}_{b_L^c}=2 y_b^2\tan^2\theta_2~,\nn\\
\Delta \tl{G}_{L_3}&=&y_\tau^2\tan^2\theta_2~,~~~~~~
\Delta \tl{G}_{E_L^c}=2 y_\tau^2\tan^2\theta_2~.
\eeqa

\item  The scalar soft parameters are given by
\beqa
m^2_{{H}_u}~~&=&\f{F_\phi^2}{16\pi^2}\[\f{3}{2}G_2\al^2_2+\f{3}{10}G_1\al^2_1\]
+\f{F_\phi^2}{(16\pi^2)^2}\[ 3y_t^2\tl{G}_{y_t}\]~,~\nn\\
m^2_{{H}_d}~~&=&\f{F_\phi^2}{16\pi^2}\[\f{3}{2}G_2\al^2_2+\f{3}{10}G_1\al^2_1\]
+\f{F_\phi^2}{(16\pi^2)^2}\[3y_b^2\tl{G}_{y_b}+y_\tau^2\tl{G}_{y_\tau}\]~,~\nn\\
m^2_{\tl{Q}_{L,a}}&=&\f{F_\phi^2}{16\pi^2}\[\f{8}{3} G_3 \al^2_3+\f{3}{2}G_2\al^2_2+\f{1}{30}G_1\al^2_1\]
+\delta_{a,3}\f{F_\phi^2}{(16\pi^2)^2}\[y_t^2\tl{G}_{y_t}+y_b^2\tl{G}_{y_b}\]
+\delta_{a,3}\Delta m^2_{\tl{Q}_{L,3}}~,\nn\\
m^2_{\tl{U}^c_{L,a}}&=&\f{F_\phi^2}{16\pi^2}\[\f{8}{3} G_3 \al^2_3+\f{8}{15}G_1\al^2_1\]
+\delta_{a,3}\f{F_\phi^2}{(16\pi^2)^2}\[2y_t^2\tl{G}_{y_t}\]~,\nn\\
m^2_{\tl{D}^c_{L;a}}&=&\f{F_\phi^2}{16\pi^2}\[\f{8}{3} G_3 \al^2_3+\f{2}{15}G_1\al^2_1\]
+\delta_{a,3}\f{F_\phi^2}{(16\pi^2)^2}\[2y_b^2\tl{G}_{y_b}\]
+\delta_{a,3}\Delta m^2_{\tl{D}^c_{L;a}}~,~\nn\\
m^2_{\tl{L}_{L;a}}&=&\f{F_\phi^2}{16\pi^2}\[\f{3}{2}G_2\al_2^2+\f{3}{10}G_1\al_1^2\]
+\delta_{a,3}\f{F_\phi^2}{(16\pi^2)^2}\[y_\tau^2\tl{G}_{y_\tau}\]
+\delta_{a,3}\Delta m^2_{\tl{L}_{L;a}}~,~\nn\\
m^2_{\tl{E}_{L;a}^c}&=&\f{F_\phi^2}{16\pi^2}\f{6}{5}G_1\al_1^2
+\delta_{a,3}\f{F_\phi^2}{(16\pi^2)^2}\[2y_\tau^2\tl{G}_{y_\tau}\]
+\delta_{a,3}\Delta m^2_{\tl{E}_{L;a}^c}~,
\eeqa
with
\beqa
G_i&=&-b_i~,~~~~~~~~~~
(b_1,b_2,b_3)=(\f{33}{5},1,-3)~,
\eeqa
and Yukawa deflection contributions
\beqa
\Delta m^2_{\tl{Q}_{L,3}}&=&\f{d^2 F_\phi^2}{(16\pi^2)^2}\[y_{Q_3 X_d \tl{b}_R }^2G^+_{Q_3 X_d \tl{b}_R}\]=\f{d^2 F_\phi^2}{(16\pi^2)^2}\[y_b^2\tan^2\theta_2G^+_{Q_3 X_d \tl{b}_R}\]~,\nn\\
\Delta m^2_{\tl{D}^c_{L;a}}&=&\f{d^2 F_\phi^2}{(16\pi^2)^2}\[2y_{Q_3 X_d \tl{b}_R }^2G^+_{Q_3 X_d \tl{b}_R}\]=\f{d^2 F_\phi^2}{(16\pi^2)^2}\[2y_b^2\tan^2\theta_2G^+_{Q_3 X_d \tl{b}_R}\]~,\nn\\
\Delta m^2_{\tl{L}_{L;a}}&=&\f{d^2 F_\phi^2}{(16\pi^2)^2}\[y_{L_3 X_d \tl{\tau}_R }^2G^+_{L_3 X_d \tl{\tau}_R }\]=\f{d^2 F_\phi^2}{(16\pi^2)^2}\[y_\tau^2\tan^2\theta_2G^+_{L_3 X_d \tl{\tau}_R }\]~,\nn\\
\Delta m^2_{\tl{E}_{L;a}^c}&=&\f{d^2 F_\phi^2}{(16\pi^2)^2}\[2y_{L_3 X_d \tl{\tau}_R }^2G^+_{L_3 X_d \tl{\tau}_R }\]=\f{d^2 F_\phi^2}{(16\pi^2)^2}\[2y_\tau^2\tan^2\theta_2G^+_{L_3 X_d \tl{\tau}_R }\]~.\nn\\
\eeqa
Here $d=-2$ and $\delta_{a,3}$ is the Kronecker delta. The beta functions for $y_{Q_3 X_u \tl{t}_R}$ and $y_{L_3 X_d \tl{\tau}_R}$ upon the messenger threshold $F_\phi$ are given by
\beqa
G_{Q_3 X_d \tl{b}_R}^+&=&y_t^2+3y_b^2+(6y_b^2+y_\tau^2)\tan^2\theta_2-\f{16}{3}g_3^2-{3}g_2^2-\f{7}{15}g_1^2~,\nn\\
G_{L_3 X_d \tl{\tau}_R}^+&=&3y_\tau^2+(3y_b^2+4y_\tau^2)\tan^2\theta_2-{3}g_2^2-\f{9}{5}g_1^2~.
\eeqa

\eit

\subsection{SUSY KSVZ axion in (deflected)AMSB}

 It will be seen soon that in the allowed parameter space of the previous SUSY spectrum,
 the lightest ordinary supersymmetric particle(LOSP) can not act as a good dark matter candidate.
 Fortunately, the axino, which is the SUSY partner of the axion to solve the strong-CP problem
 by the PQ mechanism, can act as a DM candidate if it is the true LSP\cite{kim,yamaguchi,chun,dine,baer}.

 We introduce the following prototype axion superpotential and KSVZ-type coupling involving $N_{PQ}$ species of heavy PQ messengers $Q_i,\tl{Q}_i$ in the ${\bf 5},\bar{\bf 5}$  representations of SU(5) gauge group
 \beqa
 W\supseteq \la_0 X(S\tl{S}-f^2\phi^2)+\sum\limits_{i=1}^{N_{PQ}}y_Q^i S \tl{Q}_i Q_i~,
 %%%+ y_D S \tl{D}_i D_i~, possible SU(2)_L anomaly
 %%%%+y_B \tl{S}\tl{Q}_2 Q_2~, cannot exist, otherwise no-anomaly.
 \eeqa
 with the PQ charge assignments
 \beqa
 PQ(X)=0,~PQ(S)=-PQ(\tl{S})=1~,~PQ(Q_i)=PQ(\tl{Q}_i)=-1/2.
 \eeqa
  Since the global $U(1)_{PQ}$ symmetry is anomalous under QCD, the strong CP problem can be solved.

  In the SUSY limit, the scalar potential for $X,S,\tl{S}$ after integrating out the PQ messengers can be given as
\beqa
V_0=\la^2_0|X|^2\(|S|^2+|\tl{S}|^2\)+\la_0^2|S\tl{S}-f^2|^2~.
\label{prototype}
\eeqa
The PQ scalar, however, will not be stabilized because there is a moduli space characterized by
$S\tl{S}=f^2\phi^2$ with $X=0$, which parameterize the scale transformation adjunct to the
complexified $U(1)_{PQ}$ symmetry\cite{kim1}.
This argument breaks down if we take into account the SUSY breaking effect.
Thus, in order to stabilize the PQ scalar at an appropriate scale, we have to take into account
the SUSY breaking effects in the scalar potential.
In this scenario, we will include the AMSB-type SUSY breaking effects in the potential.

We have the discontinuity of the anomalous dimension for $S$ across the PQ messenger threshold determined
by $\Lambda_Q\equiv\la_0\langle S\rangle$
\beqa
G_S^U&=&-\f{1}{8\pi^2}\[ \sum\limits_{i}5(y^i_Q)^2+\la_0^2\]~,\nn\\
\Delta G_S &=&-\f{1}{8\pi^2}\[ \sum\limits_{i}5(y^i_Q)^2\]~,
\eeqa
with $G_S^U$ the anomalous dimension of $S$ upon the $\tl{Q}_i,Q_i$ scale $\Lambda_Q$. So we can obtain that the discontinuity of $\beta_{y_Q^i},\beta_{\la_0}$ acrossing $\Lambda_Q$
\beqa
\Delta \beta_{y_Q^i}&=&\f{1}{16\pi^2}\[2(y^i_Q)^2+\sum\limits_{j}5(y^j_Q)^2+\la_0^2\]~,\nn\\
\Delta \beta_{\la_0}&=&\f{1}{16\pi^2}\[\sum\limits_{j}5(y^j_Q)^2\]~.
\eeqa
 The soft SUSY parameters for $S$ from AMSB with Yukawa deflections can be given similarly as eqn.(\ref{soft:scalar})
\beqa
m^2_{S}&=&\f{F_\phi^2}{(16\pi^2)^2}\left\{\f{}{}3\la_0^4- [(d^\pr)^2+2d^\pr]\la_0^2
\[ \sum\limits_{i}5(y^i_Q)^2\]\right.\nn\\
&&~~~~~~~~\left.\f{}{}+(d^\pr)^2\sum\limits_{i}5(y_Q^i)^2
\[2(y^i_Q)^2+\sum\limits_{j}5(y^j_Q)^2+\la_0^2\]\right\},
\eeqa
with $d^\pr$ a typical deflection parameter to characterize the deflection induced by integrating out the heavy PQ messenger fields.

The soft SUSY parameters for gauge singlets $\tl{S},X$ come entirely from AMSB, which will not receive additional Yukawa deflection contributions
\beqa
m^2_{\tl{S}}=m^2_{X}=\f{F_\phi^2}{(16\pi^2)^2}\[3\la_0^4\].
\eeqa

The form of the trilinear couplings $A_{\la_0} X S\tl{S}$ at the $\Lambda_Q$ scale will  be generated by
\beqa
A_{\la_0}=\la_0\f{F_\phi}{16\pi^2}\[3\la_0^2-d^\pr(\sum\limits_{i}5(y^i_Q)^2) \]~.
\eeqa
So the full potential for $S,\tl{S},X$ will be given by
\beqa
V(S,\tl{S},X)=m^2_{S}|S|^2+m^2_{\tl{S}}|{\tl{S}}|^2+m^2_{X}|X|^2+A_{\la_0} X S\tl{S}+2\la_0 F_\phi f^2 (X+X^\da)+V_0,
\eeqa
with $V_0$ the prototype scalar potential in eqn.(\ref{prototype}).
The minimum conditions are given by
\beqa
\[2 m^2_{X}+2\la_0^2\(v_S^2+v_{\tl{S}}^2\)\]v_X+\(4\la_0 F_\phi f^2+A_{\la_0} v_S v_{\tl{S}}\)=0~,\nn\\
\[2 m^2_{S}+2\la_0^2 v_X^2\]v_S+2\la_0^2\(v_Sv_{\tl{S}}-f^2\)v_{\tl{S}}+A_{\la_0} v_X v_{\tl{S}}=0~,\nn\\
\[2 m^2_{\tl{S}}+2\la_0^2 v_X^2\]v_{\tl{S}}+2\la_0^2\(v_Sv_{\tl{S}}-f^2\)v_{{S}}+A_{\la_0} v_X v_{{S}}=0~,
\eeqa

with
\beqa
\langle X\rangle&\equiv&v_X,~~
\langle S\rangle\equiv v_S,~~
\langle \tl{S}\rangle \equiv v_{\tl{S}}~.\nn
\eeqa

We can see that for all $\la_0,y_{Q}^i\sim {\cal O}(1)$ and $f\gg F_\phi$, the VEVs can be approximately solved to be
\beqa
v_X&\approx&\f{F_\phi}{\la_0}-\f{F_\phi m_X^2}{\la_0^3 f^2}-\f{A_{\la_0}}{4\la_0^2}~,\nn\\
v_S&\approx& f+f\f{ {m}^2_{\tl{S}}-{m}^2_{S}}{2F^2_\phi}+\f{F_\phi^2}{2\la_0^2f^2}\(1+\f{ {m}^2_{\tl{S}}+{m}^2_{S}}{F^2_\phi}\)- F_\phi\f{A_{\la_0}}{2\la^3_0f^2}~,\nn\\
v_{\tl{S}}&\approx& f-f\f{ {m}^2_{\tl{S}}-{m}^2_{S}}{2F^2_\phi}+\f{F_\phi^2}{2\la_0^2f^2}\(1+\f{ {m}^2_{\tl{S}}+{m}^2_{S}}{F^2_\phi}\)- F_\phi\f{A_{\la_0}}{2\la^3_0f^2}~.
\eeqa
In this limit, the deflection parameter $d^\pr$ can be determined to be
\beqa
d^\pr &\equiv& \f{F_{S}}{S F_\phi}-1
%%%% =-\la_0\f{v_X v_{\tl{S}}}{v_S F_\phi}-1~,\nn\\
  \approx-\la_0\f{v_X}{F_\phi}-1~\approx-2.
\eeqa

The PQ breaking scale $f_{PQ}$ can be determined by
\beqa
f_{PQ}\approx \sqrt{v_S^2+v_{\tl{S}}^2}/N_{DW}\sim f~,
\eeqa
 which is constrained to lie within the  $'axion~window'$ at $10^9 {\rm GeV}\lesssim f_{PQ}\lesssim 10^{12} {\rm GeV}$ by astrophysical and cosmological observations\cite{axion:f}. Here $N_{DW}=N_{PQ}$ is the domain wall number. The axino, which is  the fermionic components of $(S-\tl{S})/\sqrt{2}$, acquires a mass $\la_0 v_X\approx F_\phi$. So we can see that the axino will in general be heavier than the soft SUSY breaking masses predicted by (d)AMSB, which are typically of order $F_\phi/16\pi^2$. This conclusion agrees with the results in \cite{axino:AMSB} for ordinary AMSB.

After integrating out the PQ messengers, the following effective term can be generated
\beqa
-{\cal L} &\supseteq & N_{PQ}\f{\al_i}{8\pi}\int d^2\theta \ln(S) W_i^aW^{ia}+h.c.~,\nn\\
  &\supseteq& N_{PQ}\f{\al_i}{8\pi}\int d^2\theta \f{F_S}{S}\theta^2 W_i^aW^{ia}+h.c.~, \nn\\
  &=& -N_{PQ}\f{\al_i}{8\pi} F_\phi \la^a_i\la^a_i~,
\eeqa
which  will contribute to gaugino masses
\beqa
\delta M_i=-N_{PQ}\f{\al_i}{4\pi}F_\phi~.
\label{gaugino3}
\eeqa
Combining eqn.(\ref{gaugino1}) [or eqn.(\ref{gaugino2})] with eqn.(\ref{gaugino3}), the gaugino masses can be given as
\beqa
M_i=-F_\phi\f{\al_i(\mu)}{4\pi}\[b_i-(-2)(1+N_S)-(-2)\f{N_{PQ}}{2}\]~,
\eeqa
if the RGE effects between $F_\phi$ (which typically lies between $10^5$ GeV and $10^8$ GeV in AMSB) and $f_{PQ}$ are neglected. So it can be seen that ordinary messengers and PQ messengers play a similar role for the deflection contributions to the gaugino masses. Other soft SUSY breaking parameters will neither receive contributions from PQ messengers nor from ordinary messengers at the UV scale.

As noted earlier, the axino, which acquires a mass typically at $F_\phi$, is heavier than ordinary SUSY particles. However, there is a possible way to generate a light axino mass. We can add holomorphic terms for $S,\tl{S},X $ to the Kahler potential in addition to standard canonical kinetic terms
 \beqa
 K\supseteq (X^\da X+ S^\da S+\tl{S}^\da\tl{S})+(c_S \tl{S}S+c_X X^2+h.c.)~,
 \eeqa

  Following eqn.(\ref{kahler:soft-X}), the scalar mass parameters for $S$, $\tl{S}$ and $X$ will receive additional contributions from anomaly mediation
 \beqa
{\cal L}\supseteq -c_S |F_\phi|^2  \tl{S}S-c_X |F_\phi|^2 X^2 + F_\phi^\da\int d^2\theta \[c_S \tl{S}S+c_X X^2 \]+h.c.
 \eeqa
Then the scalar potential is changed into
\beqa
V(S,\tl{S},X)&=&\tl{m}^2_{S}|S|^2+\tl{m}^2_{\tl{S}}|{\tl{S}}|^2+\tl{m}^2_{X}|X|^2+c_X |F_\phi|^2  (X^2+X^{*2})+c_S |F_\phi|^2  (\tl{S}S+\tl{S}^*S^*)~,\nn\\
&+&A_{\la_0} X S\tl{S}+2\la_0 F_\phi f^2 (X+X^\da)+\la^2_0|X|^2\(|S|^2+|\tl{S}|^2\)+\la_0^2|S\tl{S}-f^2|^2,
\eeqa
with
\beqa
\tl{m}^2_{S}&=&{m}^2_{S}+c_S^2 F_\phi^2~,~~~
\tl{m}^2_{\tl{S}}={m}^2_{\tl{S}}+c_S^2 F_\phi^2~,~~~
\tl{m}^2_{X}={m}^2_{\tl{S}}+c_X^2 F_\phi^2~.
\eeqa

The minimum conditions are given by
\beqa
2\[ \tl{m}^2_{X}+\la_0^2\(v_S^2+v_{\tl{S}}^2\)+2c_X |F_\phi|^2\]v_X+\(4\la_0 F_\phi f^2+A_{\la_0} v_S v_{\tl{S}}\)=0~,\nn\\
2\[\tl{m}^2_{S}+\la_0^2 v_X^2\]v_S+2c_S |F_\phi|^2v_{\tl{S}} +2\la_0^2\(v_Sv_{\tl{S}}-f^2\)v_{\tl{S}}+A_{\la_0} v_X v_{\tl{S}}=0~,\nn\\
2\[\tl{m}^2_{\tl{S}}+\la_0^2 v_X^2\]v_{\tl{S}}+2c_S |F_\phi|^2v_{{S}}+2\la_0^2\(v_Sv_{\tl{S}}-f^2\)v_{{S}}+A_{\la_0} v_X v_{{S}}=0~,
\eeqa
with the minimum
\beqa
v_X&\approx&\f{F_\phi}{\la_0}-\f{F_\phi  (\tl{m}^2_{X}+2c_X |F_\phi|^2)}{\la_0^3 f^2}-\f{A_{\la_0}}{4\la_0^2}~,\nn\\
v_S&\approx& f+f\f{\tl {m}^2_{\tl{S}}-\tl{m}^2_{S}}{2F^2_\phi}+\f{F_\phi^2}{2\la_0^2f^2}\(1+\f{ \tl{m}^2_{\tl{S}}+\tl{m}^2_{S}}{F^2_\phi}\)- F_\phi\f{A_{\la_0}+2c_S |F_\phi|^2}{2\la^3_0f^2}~,\nn\\
v_{\tl{S}}&\approx& f-f\f{\tl {m}^2_{\tl{S}}-\tl{m}^2_{S}}{2F^2_\phi}+\f{F_\phi^2}{2\la_0^2f^2}\(1+\f{ \tl{m}^2_{\tl{S}}+\tl{m}^2_{S}}{F^2_\phi}\)- F_\phi\f{A_{\la_0}+2c_S |F_\phi|^2}{2\la^3_0f^2}~.
\eeqa
The axino mass are therefore given by
\beqa
m_{\tl{a}}&=&\la_0 v_X +c_S F_\phi^\da~,\nn\\
&\approx& F_\phi-\f{F_\phi  (\tl{m}^2_{X}+2c_X |F_\phi|^2)}{\la_0^2 f^2}-\f{A_{\la_0}}{4\la_0}+c_S F_\phi~,
\eeqa
which can be much lighter than $F_\phi$ for $c_S\approx-1$. So the axino can possibly be the LSP and act as the DM candidate.
\subsection{The $\mu-B\mu$ problem}
In AMSB, the generation of $\mu-B\mu$ term is always troublesome because of the constraints from EWSB.
It was argued that the following holomorphic term, \beqa
\int d^4\theta \f{\phi^\da}{\phi}c_b H_u H_d~,
\label{GM}
\eeqa
which possibly be present in eqn.(\ref{eigen:ab}), will lead to a too large $B\mu$ term. However, if the following $\mu$-type term is also present in the superpotential, the resulting $\mu-B\mu$ term can possibly be consistent with the EWSB condition which typically requires $B\mu\lesssim \mu^2$. In fact, the ordinary $\mu$-term in the superpotential in AMSB will receive dependence on the compensator field
 \beqa
 W&\supseteq& \mu_0\phi \tl{H}_u \tl{H}_d~,\nn\\
  &=& \mu_0\phi \(X_u\sin\theta_1+\cos\theta_1{H}_u \)\(X_d\sin\theta_2+\cos\theta_2{H}_d \)~.
 \eeqa
It will change into
 \beqa
 W\supseteq \mu_0\phi \cos\theta_1\cos\theta_2 H_u H_d~,
 \eeqa
 after integrating out the heavy messenger fields. Combining with the eqn.(\ref{GM}), we will obtain
 \beqa
 \mu&=&\mu_0\cos\theta_1\cos\theta_2+c_b F_\phi~,\nn\\
 B\mu&=&\mu_0\cos\theta_1\cos\theta_2F_\phi-c_b F^2_\phi~.
 \eeqa
An important observation is that a minus sign appears within the RHS of $B_\mu$. For
\beqa
\left|\f{\mu_0\cos\theta_1\cos\theta_2-c_b F_\phi}{c_bF_\phi}\right|\lesssim c_b~,
\eeqa
 we can obtain $B\mu\lesssim \mu^2$ with order $1/c_b$ fine tuning.
 The EWSB condition
 \beqa
 \f{M_Z^2}{2}&=&\f{m^2_{H_d}-m_{H_u}^2\tan\beta^2}{\tan^2\beta-1}-\mu^2~,~~
  \eeqa
 requires $M_Z\lesssim \mu\approx 2c_b F_\phi$, so the value of $c_b$ should satisfy
 \beqa
 c_b\sim \f{1}{16\pi^2}~,
 \eeqa
 for generic value of $m_{H_u}^2$ in (d)AMSB.

 Csaki et al\cite{EWSB1} found the other interesting possibility for EWSB condition which requires
 \beqa
 \mu^2\sim m_{H_u}^2\ll B\mu\ll m_{H_d}^2.
 \eeqa
Spectrum of this type can be realized by introducing other types of messenger-matter mixing (for example, the lepton-messenger mixing) so as that the $H_d$ soft masses can receive additional contributions from new Yukawa couplings while $H_u$ not. Such a scenario can not only generate positive slepton masses easily, but also solve the $\mu-B\mu$ problem.

The solution of $\mu-B\mu$ problem is quite model dependent. So we leave $\mu,B\mu$ as free parameters in our numerical studies with their values determined (iteratively) by EWSB conditions.

\section{\label{secIV}Numerical Results}

There are only four free parameters in each scenario, namely
  \beqa
 F_\phi, a, 0<\tan\theta_{1,2}<50~, \tan\beta~,
  \eeqa
 with $a\equiv N_S+N_{PQ}/2$ to replace the $N_S$ in eqn.(\ref{gaugino1}) and eqn.(\ref{gaugino2}). This setting do not distinguish between PQ messengers and ordinary messengers. The tiny RGE effects between $F_\phi$ and $f_{PQ}$ are neglected.

 In our scan, we require that the tachyonic slepton problem which bothers ordinary AMSB should be solved. Besides, we impose the following constraints
\bit
\item (I) The conservative lower bounds on SUSY particles by LHC\cite{CMSSM1,CMSSM2} and LEP\cite{LEP} as well as electroweak precision observables\cite{precision} from LEP:
    \bit
    \item  Gluino mass: $m_{\tl{g}} \gtrsim 1.8$ TeV .
    \item  Light stop mass: $m_{\tl{t}_1} \gtrsim 0.85$ TeV .
    \item  Light sbottom mass $m_{\tl{b}_1} \gtrsim 0.84$ TeV.
    \item  Degenerated first two generation squarks $m_{\tl{q}} \gtrsim 1.0 \sim 1.4$ TeV.
    \item  $m_{\tl{\chi}^\pm}> 103.5 {\rm GeV}$ and the invisible decay width $\Gamma(Z\ra \tl{\chi}_0\tl{\chi}_0)<1.71~{\rm MeV}$.
    \eit

\item  (II) The lightest CP-even scalar should lie in the combined mass range for the Higgs boson: $123{\rm GeV}<M_h <127 {\rm GeV}$.

\item (III)  Flavor constraints \cite{B-physics} from B-meson rare decays are imposed as
  \beqa
  &&1.7\tm 10^{-9} < Br(B_s\ra \mu^+ \mu^-) < 4.5\tm 10^{-9}~,\\
  && 0.85\tm 10^{-4} < Br(B^+\ra \tau^+\nu) < 2.89\tm 10^{-4}~,\\
  && 2.99\tm 10^{-4} < Br(B_S\ra X_s \gamma) < 3.87\tm 10^{-4}~.
   \eeqa

\item (IV) The relic density of the dark matter should satisfy the upper bound of the Planck data $\Omega_{DM}h^2 = 0.1199\pm 0.0027$ \cite{Planck} in combination with the WMAP data \cite{WMAP}(with a $10\%$ theoretical uncertainty). In our scenario, the neutralino or axino can be the DM paticle.
    The axino DM can be generated dominantly from the decay of lightest ordinary supersymmetric particle (LOSP), such as $\tl{\tau}_1,\tl{e}_R$. The left-handed sneutrino DM scenario had already been ruled out by DM direct detection experiments\cite{LUX2016,PANDAX,XENON1T2018}, so $\tl{\nu}_{eL},\tl{\nu}_{\tau L}$ etc are not good DM candidates. However, the left-handed sneutrino can possibly act as the LOSP and decay into LSP axino after it was produced in the early universe or at the collider.
\eit
We have the following numerical discussions:
\subsection*{Scenario I:}
\bit
\item  Many points can survive the constraints from (I)-(III) for $a\geq 2$. However, we check that no point can  survive the previous constraints for $a=0$ or 1. It is interesting to note that tachyonic slepton problem can not be solved for $N<5$ messenger species in ordinary Kahler deflection\cite{Nelson:2002sa} of AMSB. With Yukawa deflection induced by messenger-Higgs mixing, $3\leq 1+a<5$ messenger species are adequate to push the negative squared masses for sleptons to positive values in our scenario.

\begin{figure}[htb]
\begin{center}
\includegraphics[width=2.9in]{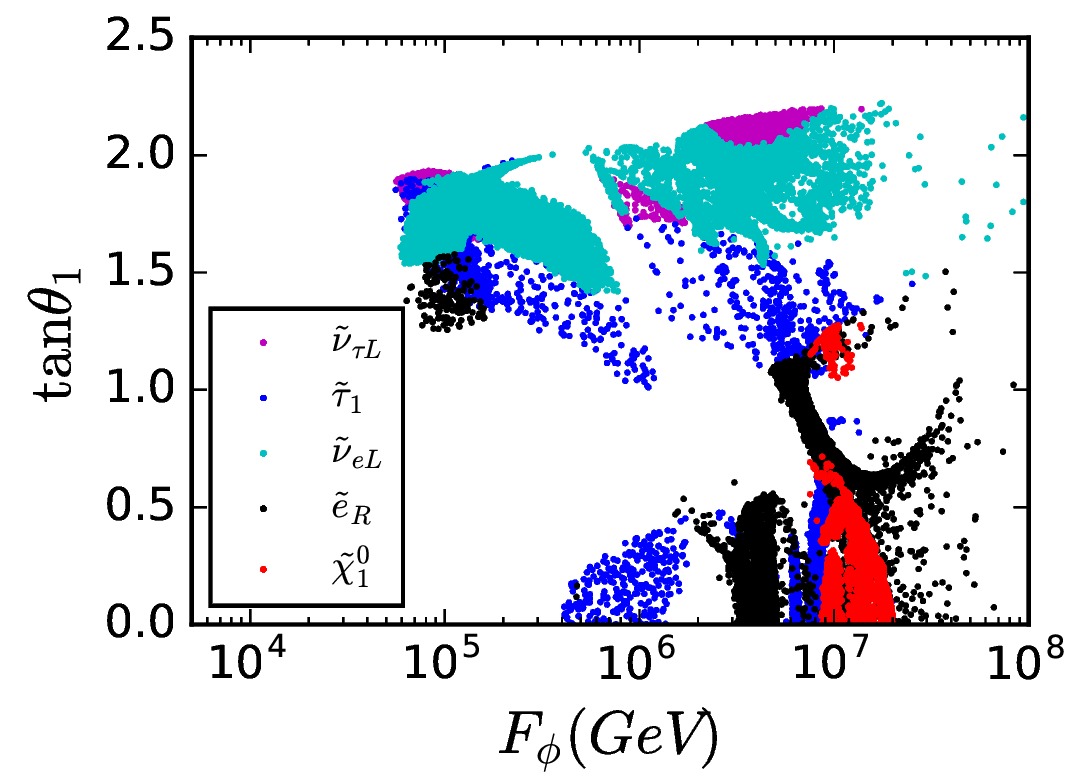}
\includegraphics[width=2.9in]{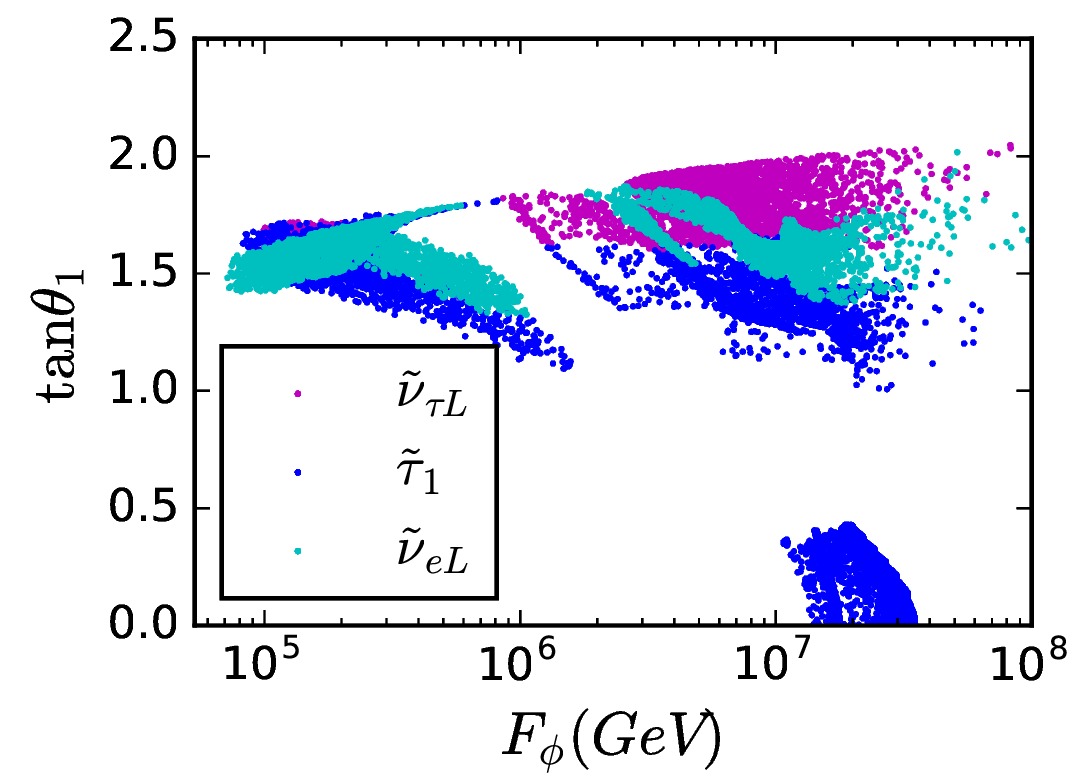}\\
\end{center}
\vspace{-.5cm}
\caption{ Allowed regions of $\tan\theta_1$ vs $F_\phi$ with $a=3$ (left panel) and $a=2$ (right panel) in scenario I. All points satisfy the constraints from (I) to (III).}
\label{fig1}
\end{figure}

     We show the allowed region of $\tan\theta_1$ versus $F_\phi$ in figure \ref{fig1}, within which various types of the LOSP are marked by various colors.
    For $a=3$, the lightest neutralino $\tl{\chi}_1^0$ can possibly be the LOSP with $F_\phi\sim10^7 {\rm GeV}$. However, for $a=2$, the lightest neutralino $\tl{\chi}_1^0$ cannot be the LOSP in the whole parameter space. Other types of superpartner, such as $\tl{\nu}_{eL},\tl{e}_R, \tl{\tau}_1$, can also serve as LOSP.

%%%%%%%%%%%%%%%%%%%%%%%%%%Fig1%%%%%%%%%%%%%%%%%%
%%%%%%%%%%%%%%%%%%%%%%%%%%%%%%%%%%%%%%%%%%%%%%%%%%%%%
 %%%%%%%%%%%%%%%%%%%%%%%%%%Fig1%%%%%%%%%%%%%%%%%%
\begin{figure}[htb]
\begin{center}
\includegraphics[width=2.9in]{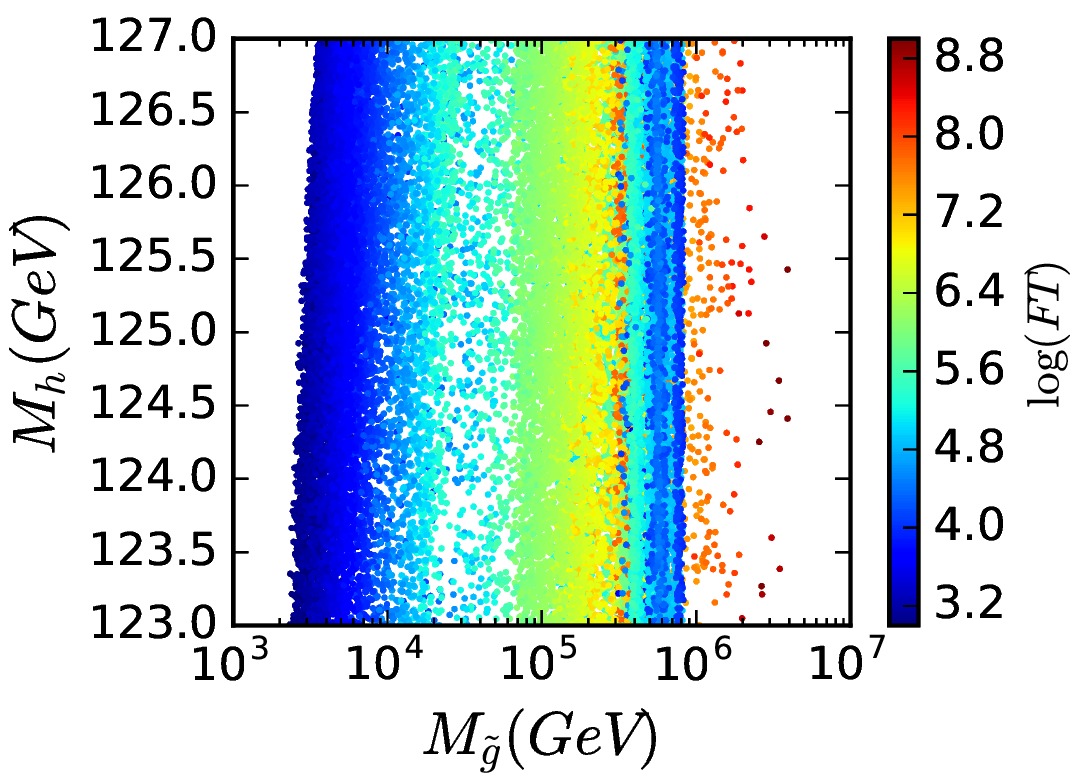}
\includegraphics[width=2.9in]{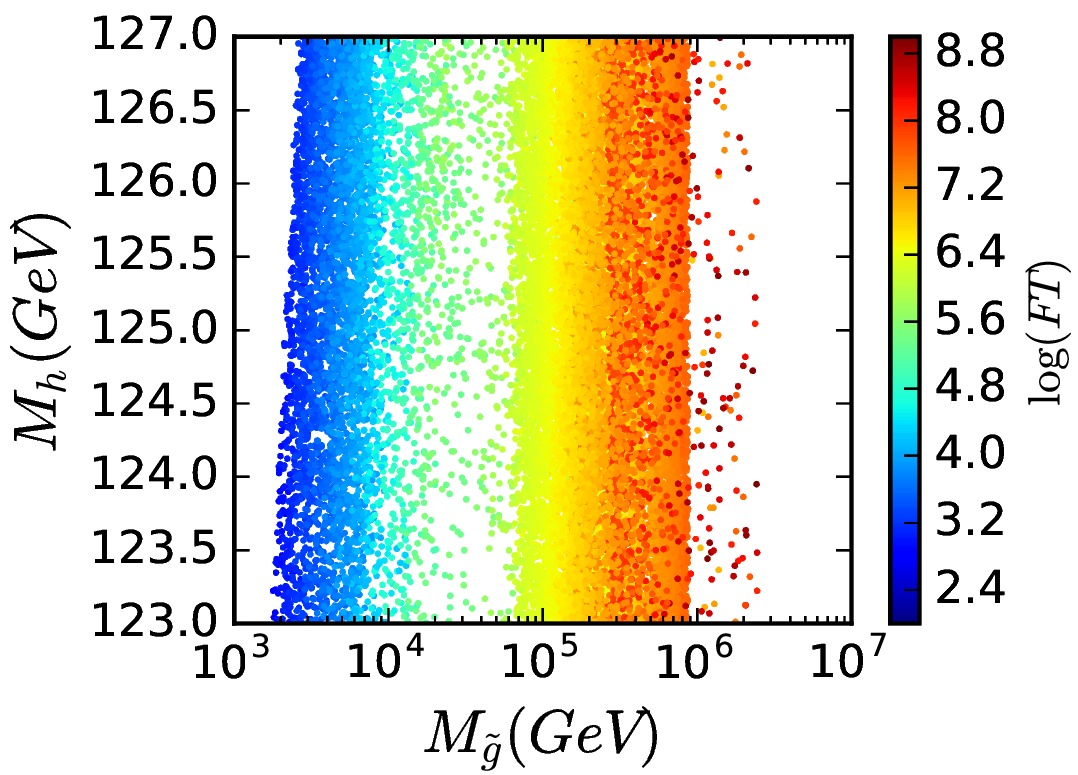}\\
\includegraphics[width=2.9in]{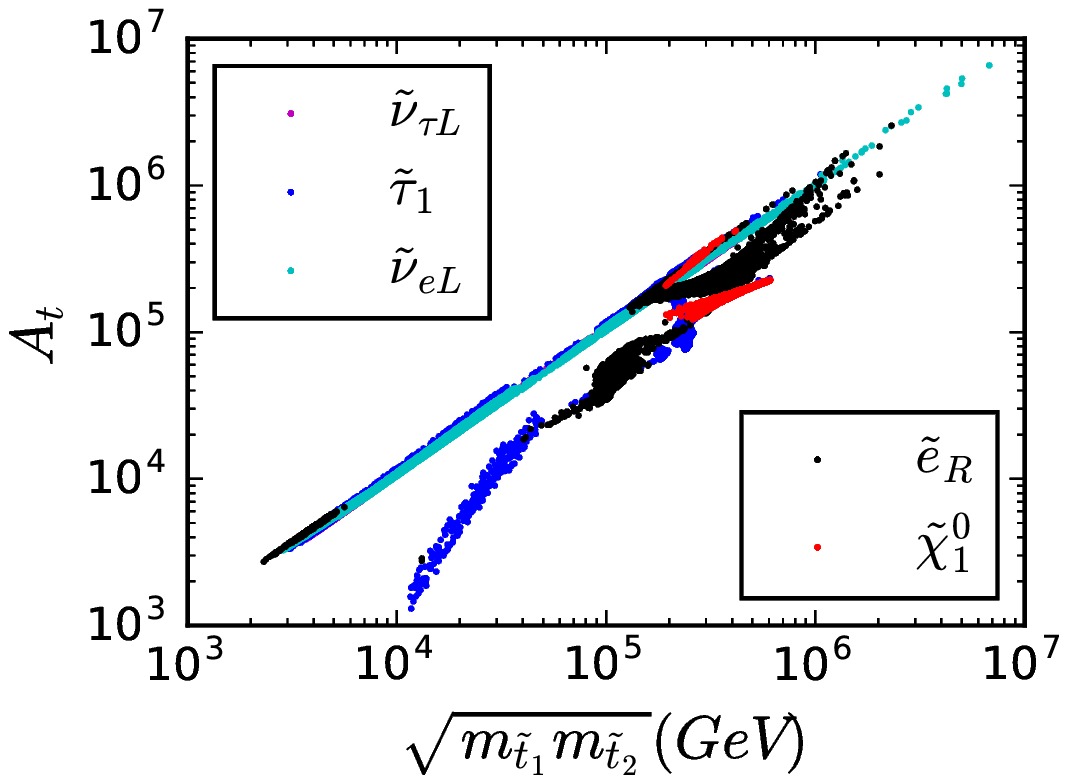}
\includegraphics[width=2.9in]{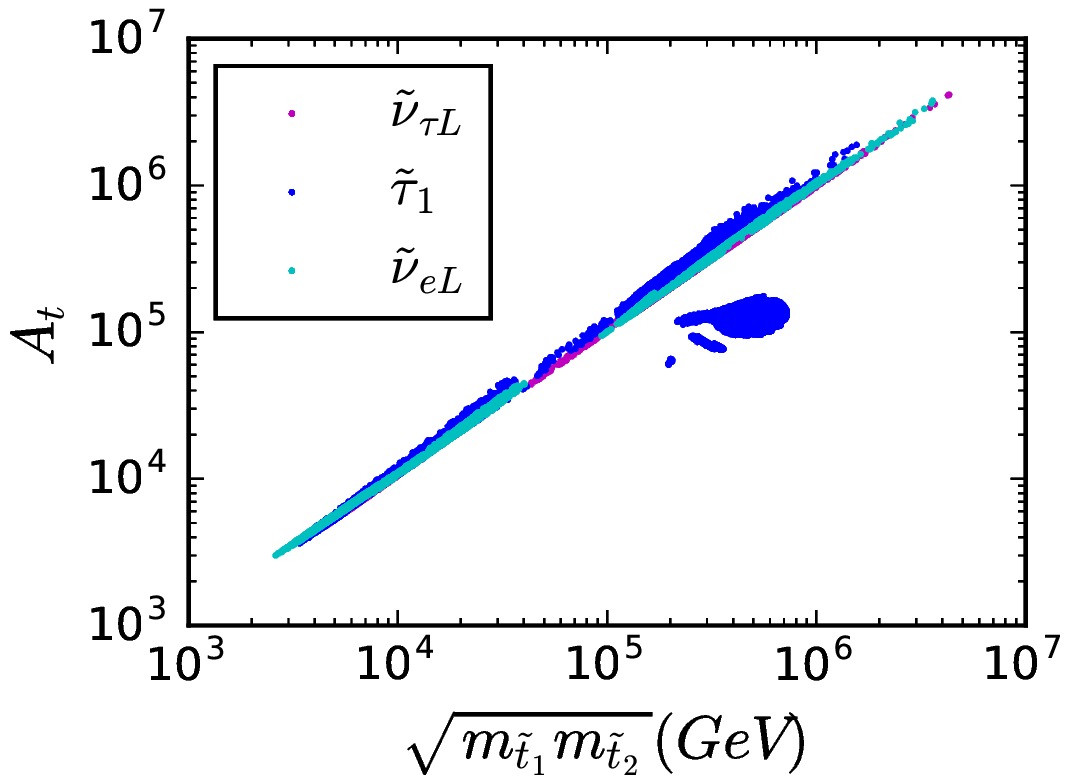}\\
\includegraphics[width=2.9in]{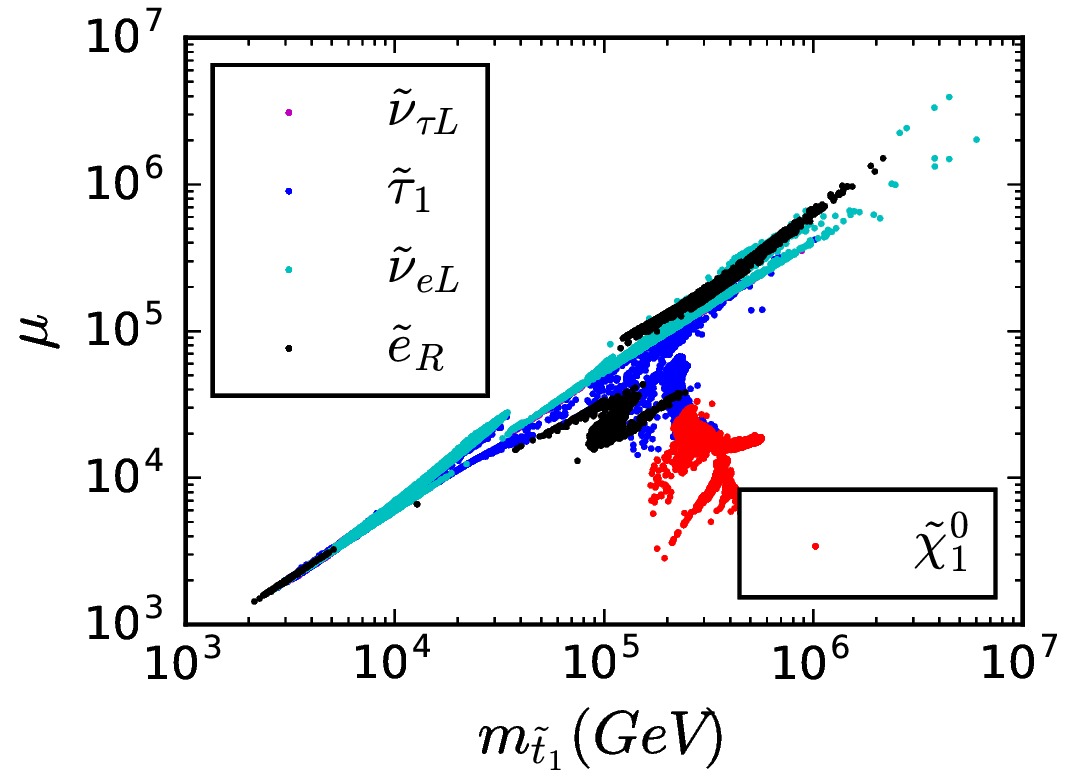}
\includegraphics[width=2.9in]{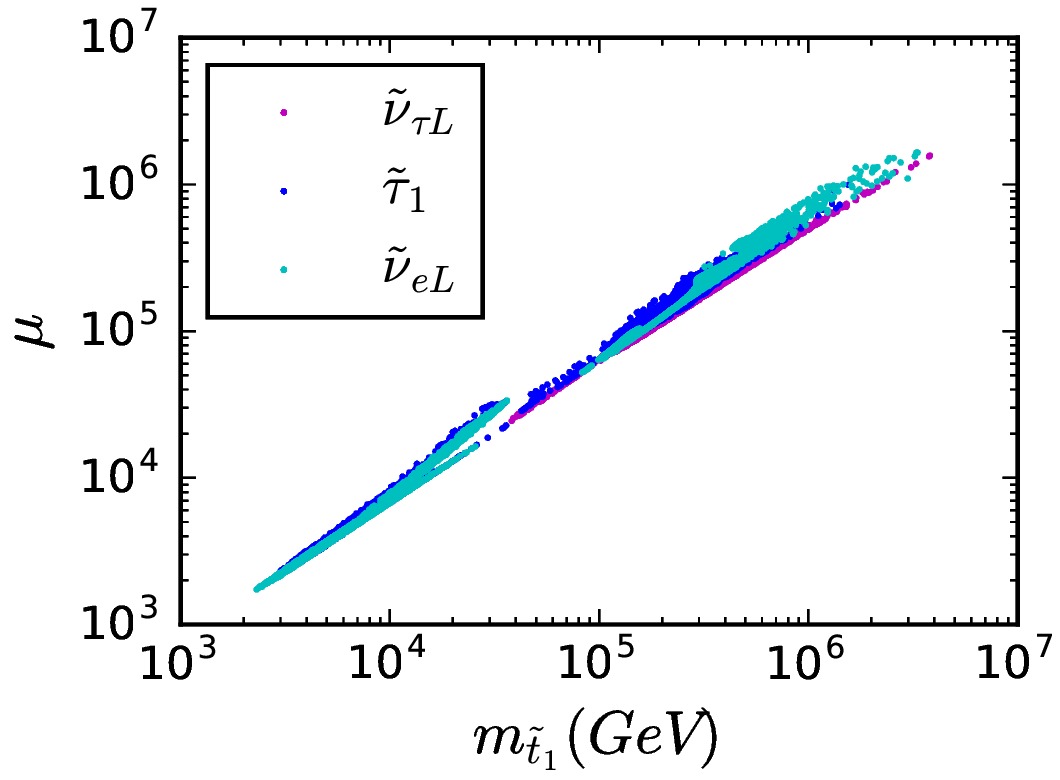}\\
\end{center}
\vspace{-.5cm}
\caption{ Allowed regions for various LOSP with $a=3$ (left panel) and $a=2$ (right panel) in scenario I. All points satisfy the constraints from (I) to (III). In the upper panels, the BGFT measure is used to parameterize the level of EWFT.}
\label{fig2}
\end{figure}
%%%%%%%%%%%%%%%%%%%%%%%%%%%%%%%%%%%%%%%%%%%%%%%%%%%%%
\item  The Higgs mass in MSSM is given by
    \beqa
     m_{h}^{2}\simeq m_{Z}^{2}\cos^{2}2\beta
      +\frac{3m_{t}^{4}}{4\pi^{2}v^{2}}
      \left[\log\frac{M_{\mathrm{SUSY}}^{2}}{m_{t}^{2}}+\frac{\tilde{A}_{t}^{2}}{M_{\mathrm{SUSY}}^{2}}\left(1-\frac{\tilde{A}_{t}^{2}}{12M_{\mathrm{SUSY}}^{2}}\right)\right],
    \eeqa
     with $\tilde{A}_{t} \equiv A_t-\mu\cot\beta$ and $M_{\mathrm{SUSY}}=\sqrt{m_{\tilde{t}_{1}}m_{\tilde{t}_{2}}}$ the geometric mean of stop masses. To increase the loop contributions to the Higgs mass, we can either choose $M_{\mathrm{SUSY}}/m_{t}\gg1$ or  $M_{\mathrm{SUSY}}/m_{t}>1$ with $\tilde{A}_{t}/M_{\mathrm{SUSY}}>1$. Without stop mixing, the stop masses have to be heavier than $5$ TeV.

     The Higgs mass $m_h$ versus the gluino mass $m_{\tl{g}}$ for the survived points are shown in the upper panels of figure \ref{fig2}. We also show the parameters $A_t$ vs $\sqrt{m_{\tl{t}_1}m_{\tl{t}_2}}$ in the middle panels of figure \ref{fig2}, which can be used to estimate the dominant loop contributions to the Higgs mass. We can see from the figures that it is fairly easy to accommodate the 125 GeV Higgs mass in our scenarios. As a large trilinear coupling $A_t$ at the messenger scale can be generated by eqn.(\ref{at1}) and eqn.(\ref{at2}), our scenario can accommodate the 125 GeV Higgs mass with the geometric mean of stop masses as low as 2 TeV.  This is in contrast to ordinary GMSB scenario, which predicts a vanishing $A_t$ at the messenger scale and is difficult to accommodate the 125 GeV Higgs mass with such light stop masses (unless the messenger scale in GMSB is extremely high).

     Low value of $F_\phi$, which sets the whole soft SUSY spectrum including the stop masses to be light, needs low electroweak fine-tuning(EWFT). The involved  Barbier-Giudice(BG) FT measures\cite{BGFT} are shown with different colors. In our sceanrio, the least BGFT value can be ${\cal O}(10^3)$. To see more clearly the EWFT, we plot the parameter $\mu$ vs $m_{\tl{t}_1}$ in the bottom panels of figure \ref{fig2}. Low EWFT in general corresponds to low value of $\mu$.

 \item  As noted previously, the LOSP in our scenarios can be the $\tl{\nu}_{eL},\tl{e}_R, \tl{\tau}_1$ other than the lightest neutralino $\tl{\chi}_1^0$. If the lightest neutralino is lighter than the axino, the $\chi_1^0$ LSP can act as the DM candidate.
     On the other hand, if axino is the LSP and act as the DM particle, the LOSP can later decay into axino after its freezing out. The relic density of axino is therefore related to that of LOSP by
     \beqa
     \Omega_{\tl{a}} h^2=\f{m_{\tl{a}}}{m_{LOSP}}\Omega_{LOSP} h^2~.
     \label{relicLOSP}
     \eeqa
    %%%%%%%%%%%%%%%%%%%%%%%%%%Fig1%%%%%%%%%%%%%%%%%%
\begin{figure}[htb]
\begin{center}
\includegraphics[width=2.9in]{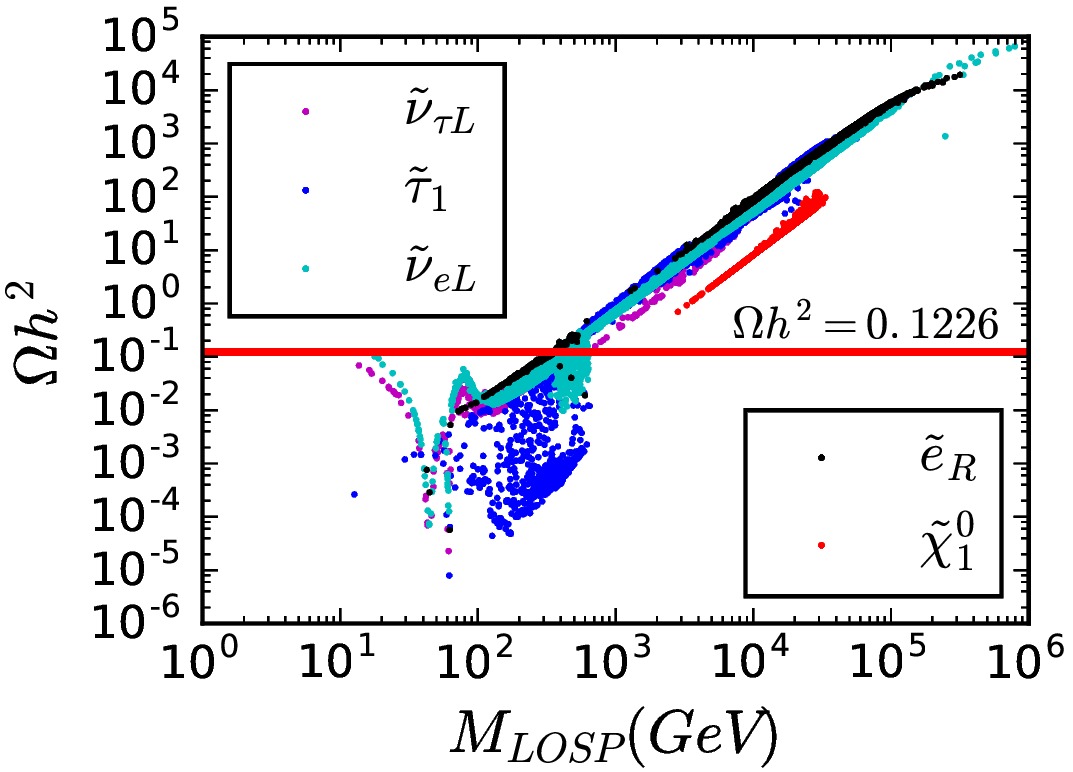}
\includegraphics[width=2.9in]{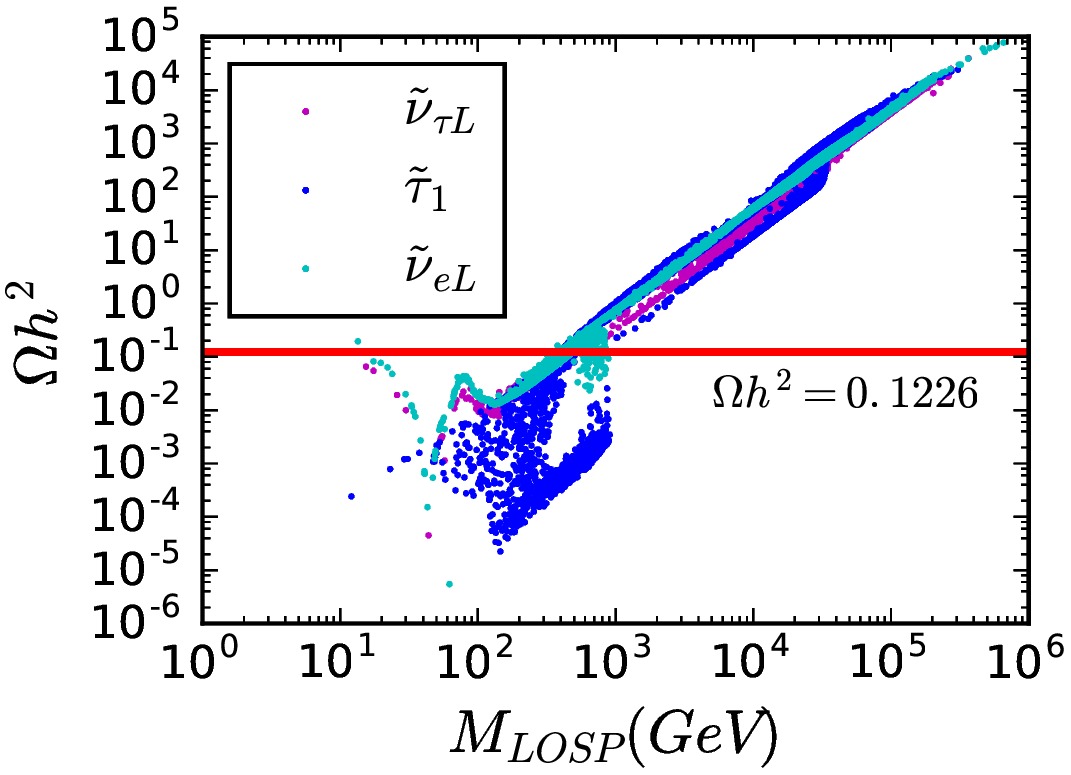}\\
\end{center}
\vspace{-.5cm}
\caption{ The relic abundances of various LOSP particles for $a=3$ (left panel) and $a=2$ (right panel) in scenario I.}
\label{fig3}
\end{figure}
%%%%%%%%%%%%%%%%%%%%%%%%%%%%%%%%%%%%%%%%%%%%%%%%%%%%%
     The relic abundances of those various LOSP are shown in figure \ref{fig3}. We can see from the figure that the lightest neutralino can serve as the LOSP for $a=3$. However, $\chi_1^0$ particle, if it is also the LSP, has a relic abundance exceeding the DM upper bound and is therefore ruled out as the DM particle. Axino DM scenario, on the other hand, is still allowed.
     It can be seen from equation (\ref{relicLOSP}) that the LSP relic abundance is always smaller than that of the LOSP. So, if axino is the LSP, the $\chi_1^0$ LOSP can decay into the axino and its relic density can therefore possibly lead to a right amount of axino DM. Other LOSP species, such as $\tl{e}_R, \tl{\tau}_1$, can not be the DM candidates because they are not electric neutral. The left-handed sneutrino DM scenario had already be rule out by DM direct detection experiments. All of these LOSPs can decay into axino DM particle after they freeze out if the axino is the true LSP.

   It is hopeless to detect the axino DM via DM direct detection experiments and collider experiments because of its extremely weak interaction strength.
   However, the axino DM may show up its existence from the properties of the LOSP.
    The LOSP typically decays into axino with a lifetime less than one second and practically be stable inside the collider detector. The electrically charged particle would appear as a stable particle inside the detector. The injection of high-energetic hadronic and electromagnetic particles, produced from late decays of the LOSP into axino (with lifetime less than one second), will not affect the abundance of light elements produced in the Big Bang Nucleosynthesis(BBN) era.
\eit
\subsection*{Scenario II:}

\begin{figure}[htb]
\begin{center}
\includegraphics[width=2.9in]{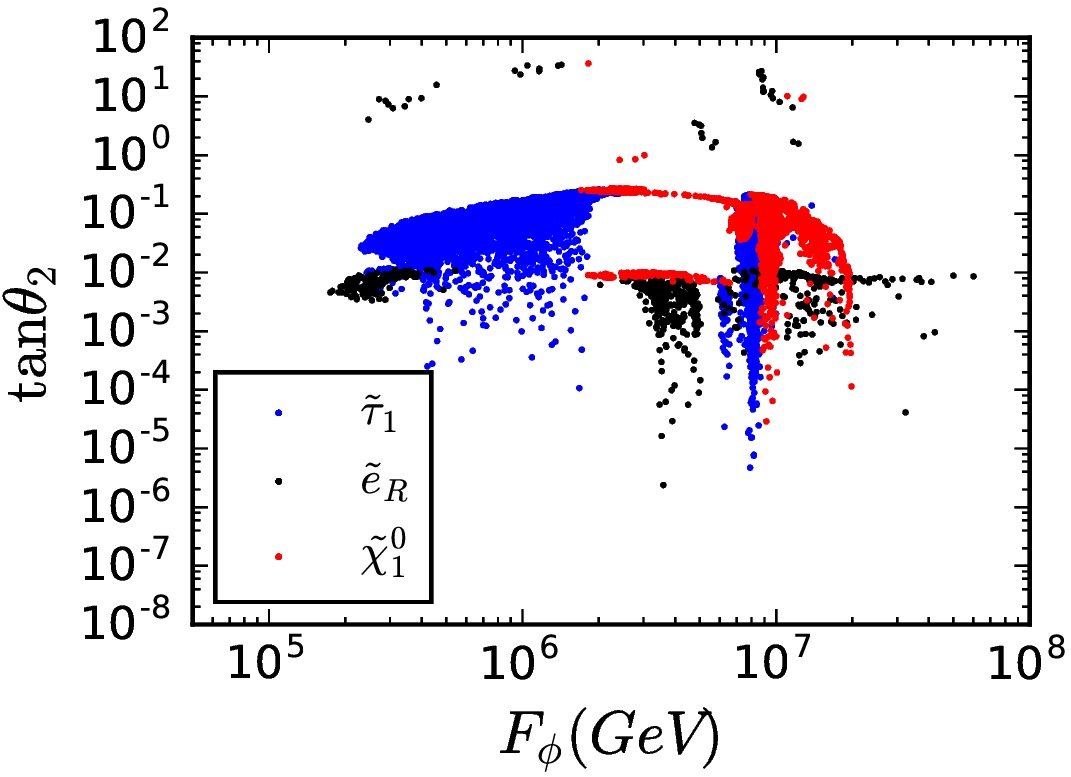}
\includegraphics[width=2.9in]{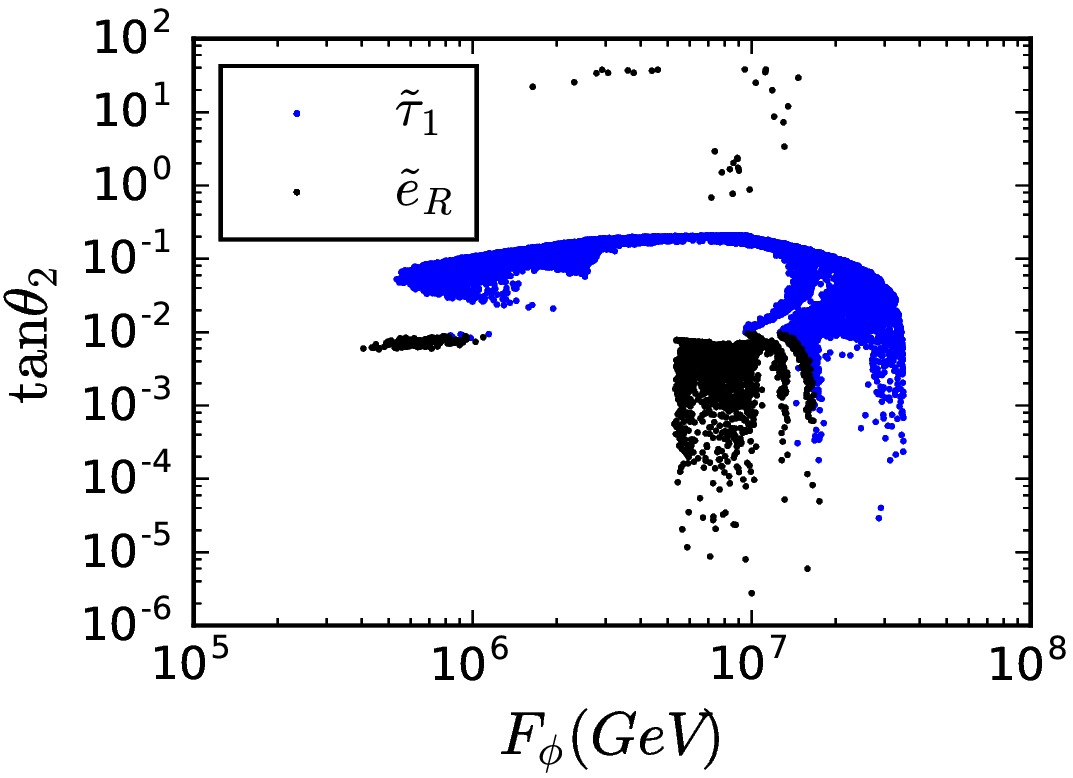}\\
\end{center}
\vspace{-.5cm}
\caption{ Allowed regions of $\tan\theta_2$ vs $F_\phi$ with $a=3$(left panel) and $a=2$(right panel) in scenario II. All points satisfy the constraints from (I) to (III).}
\label{fig4}
\end{figure}

\begin{figure}[htb]
\begin{center}
\includegraphics[width=2.9in]{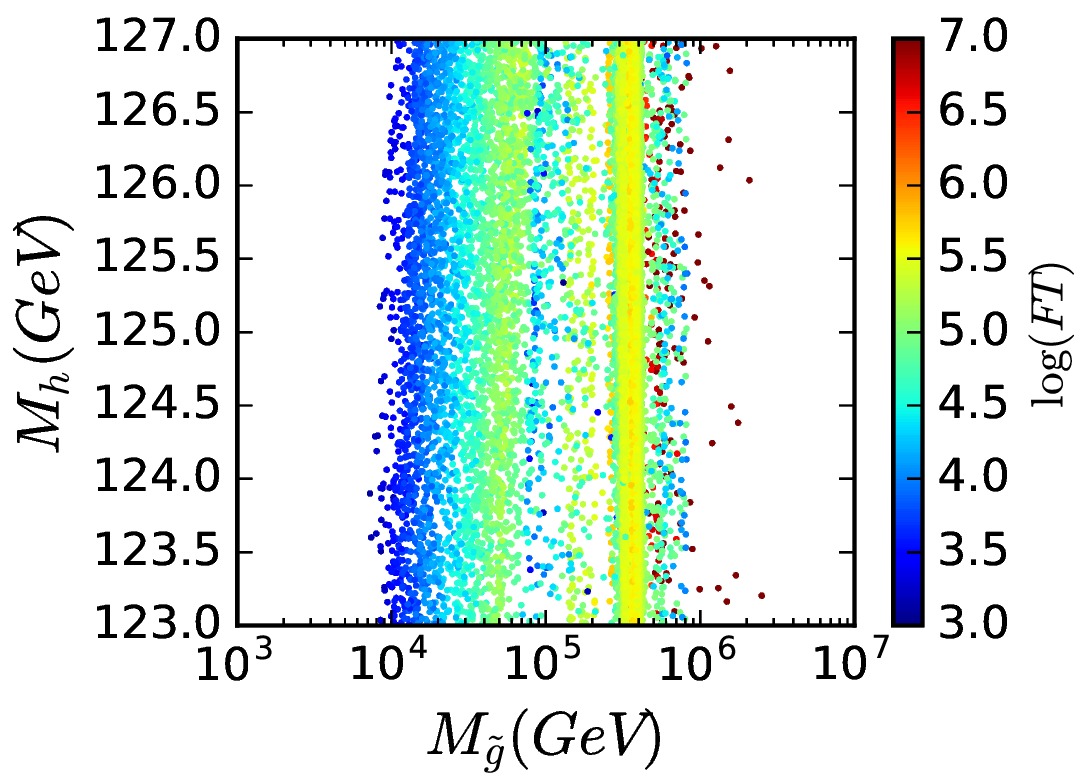}
\includegraphics[width=2.9in]{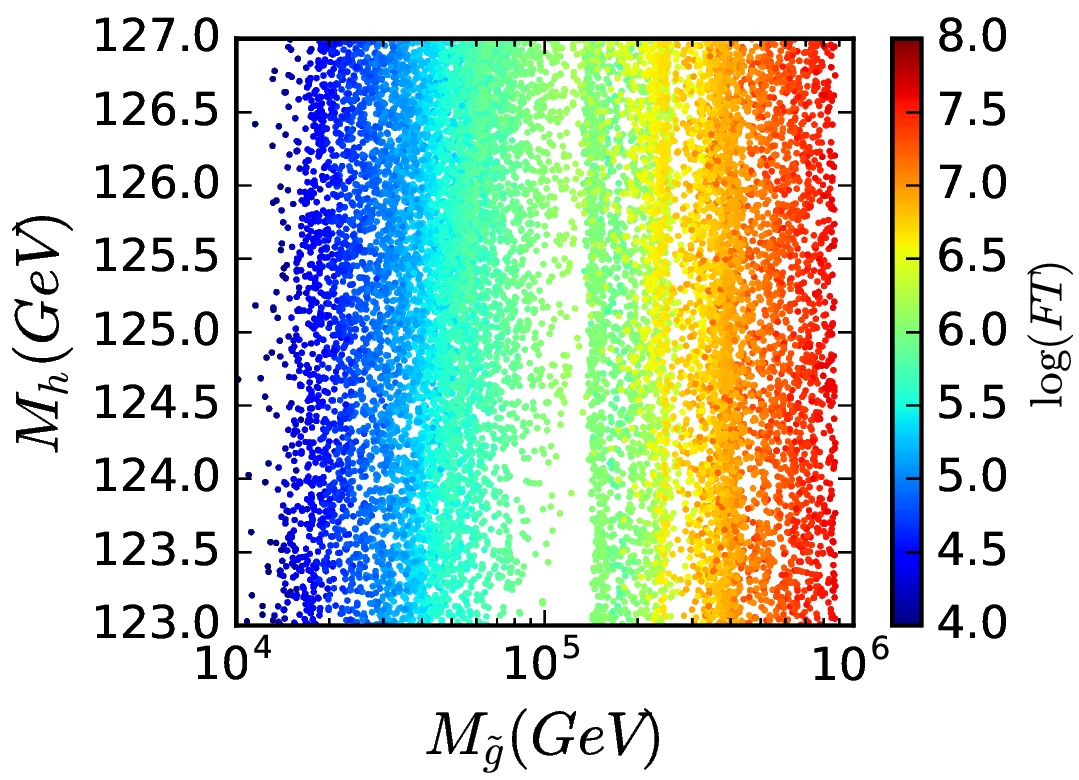}\\
\includegraphics[width=2.9in]{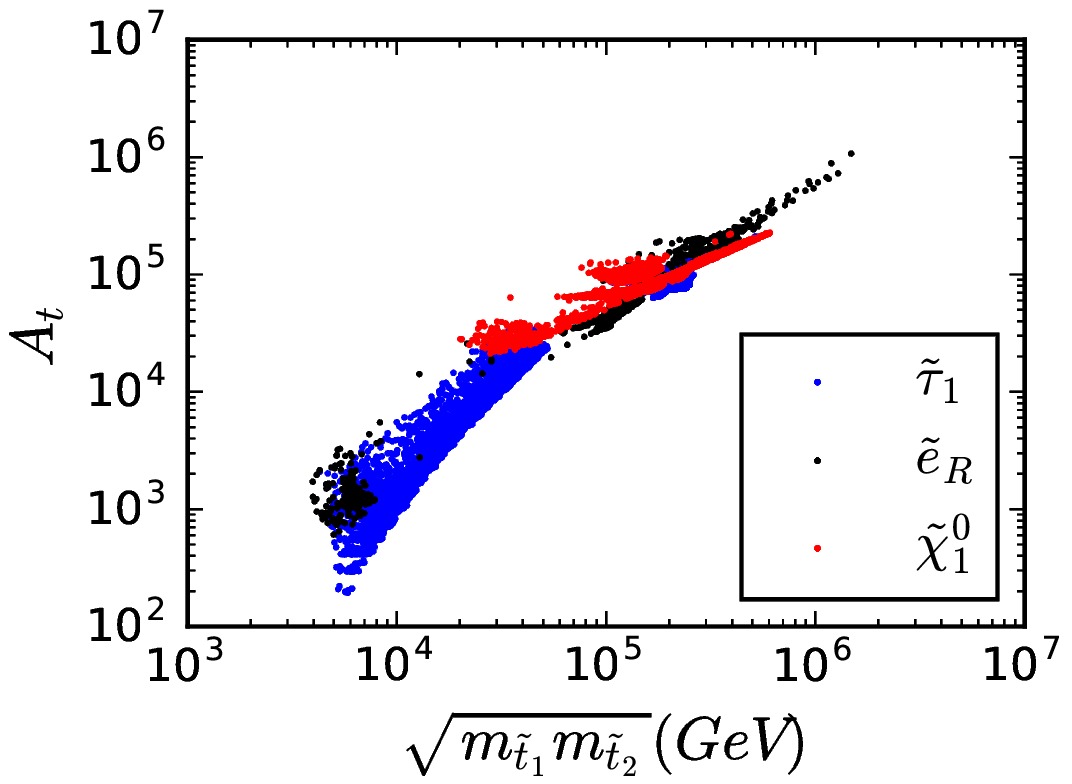}
\includegraphics[width=2.9in]{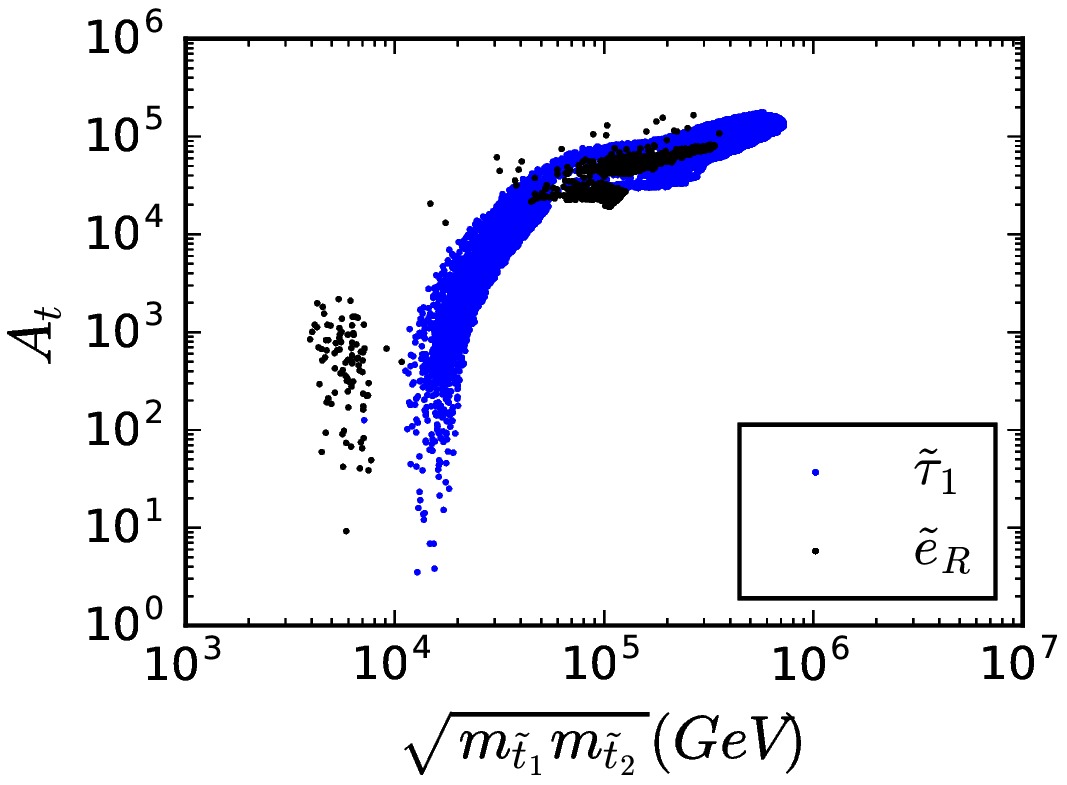}\\
\includegraphics[width=2.9in]{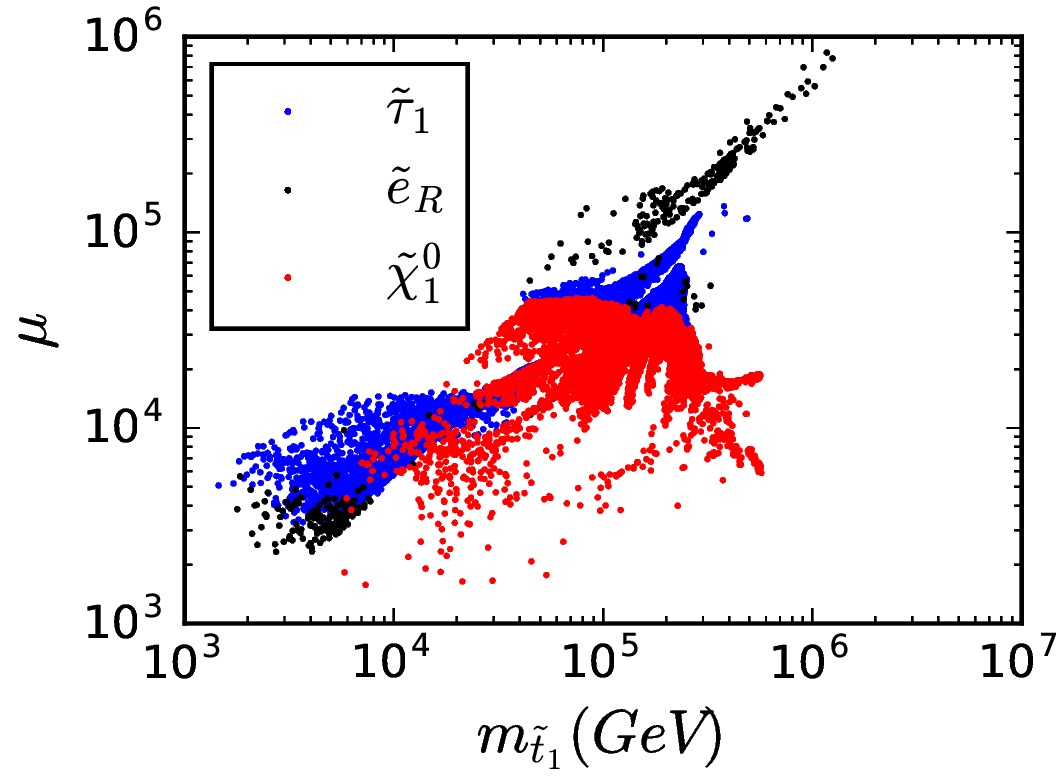}
\includegraphics[width=2.9in]{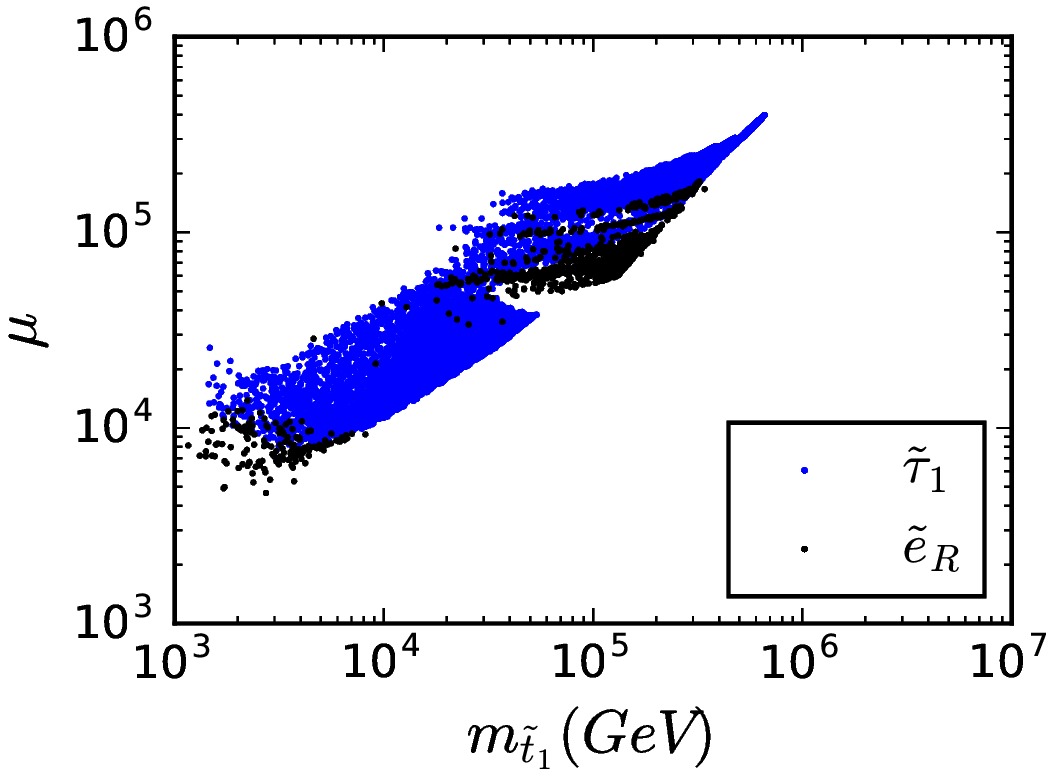}\\
\end{center}
\vspace{-.5cm}
\caption{ Allowed regions for various LOSP with $a=3$ (left panel) and $a=2$ (right panel) in scenario II. All points satisfy the constraints from (I) to (III). In the upper panels, the BGFT measure is used to parameterize the level of EWFT.}
\label{fig5}
\end{figure}

Similar discussions can be carry out for Scenario II.  Allowed regions of $\tan\theta_2$ versus $F_\phi$ for various types of the LOSP are marked with various colors in figure \ref{fig4}. As scenario I, the survived regions admit $\tl{\nu}_{eL},\tl{e}_R, \tl{\tau}_1, \chi_1^0$  as the LOSP. Besides, the 125 GeV Higgs can also be accommodated easily in this scenario. In fact, as can be seen in the middle panels of figure \ref{fig5}, $\sqrt{m_{\tilde{t}_{1}}m_{\tilde{t}_{2}}}$ can be as low as 3 TeV with an intermediate large value of $A_t$. From the allowed ranges of the $\mu$ vs $m_{\tl{t}_1}$ parameters, it is clear that the case $a=3$ can adopt relatively light $\mu$ in compare with the case $a=2$, therefore less EWFT. This observation is consistent with the conclusion from the values of the BGFT measure in the upper panels of figure \ref{fig5}.

\begin{figure}[htb]
\begin{center}
\includegraphics[width=2.9in]{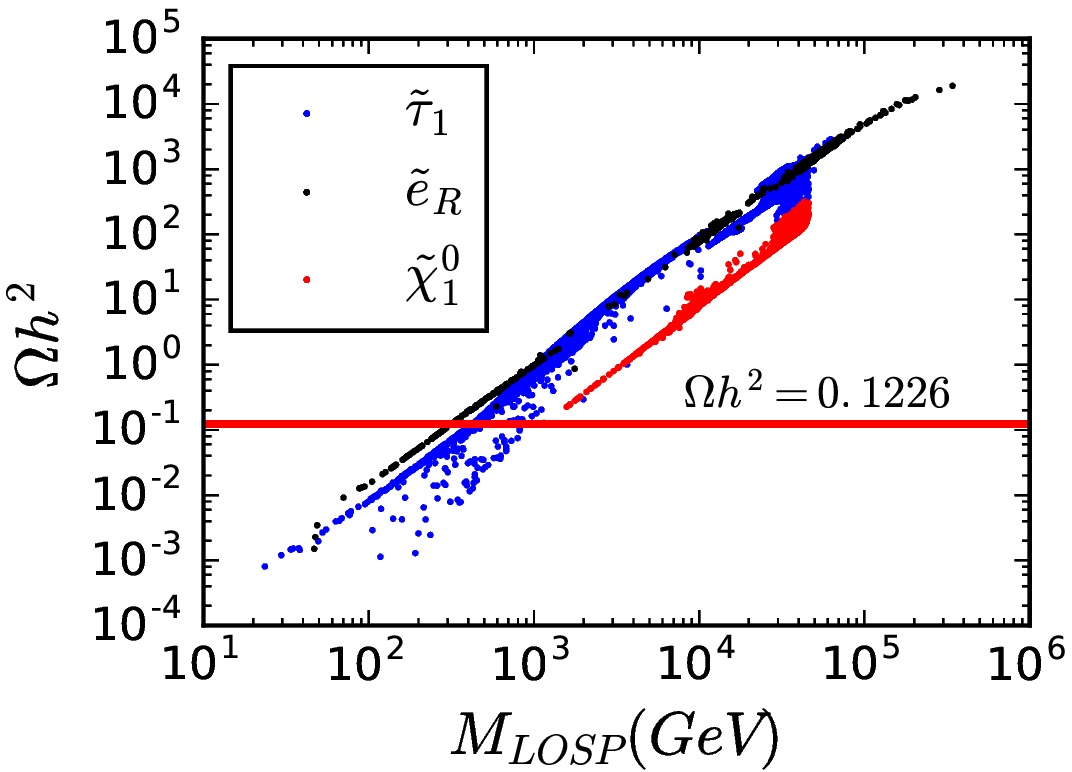}
\includegraphics[width=2.9in]{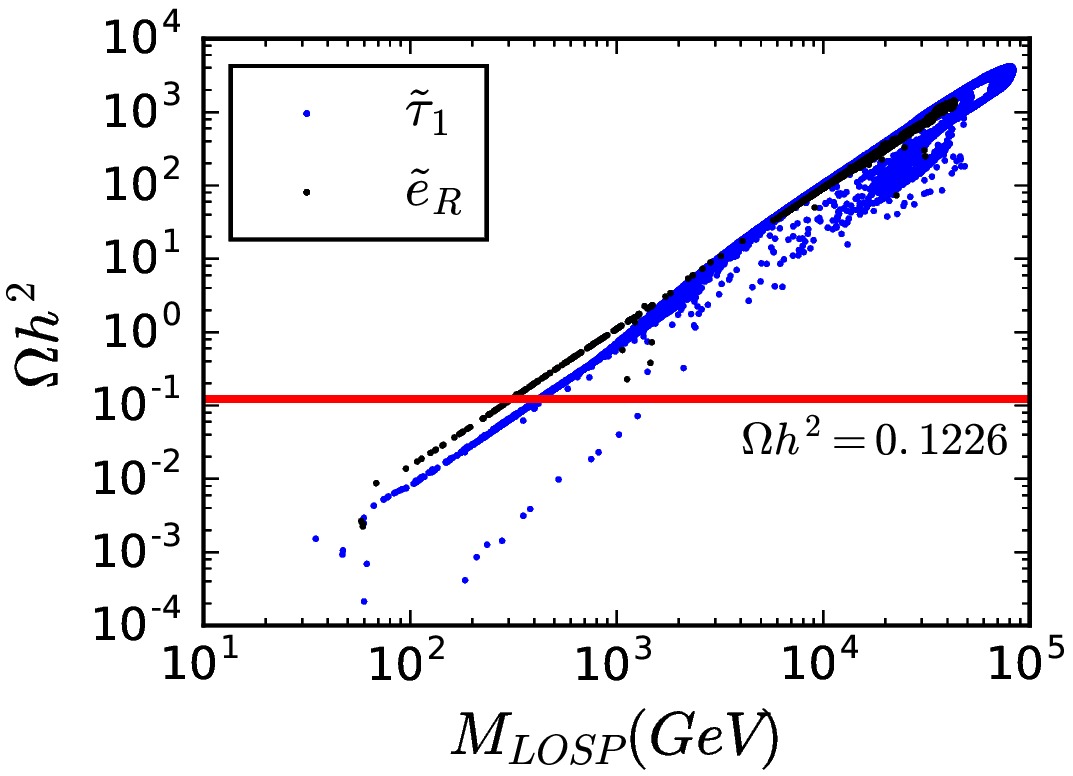}\\
\end{center}
\vspace{-.5cm}
\caption{ The relic abundances of various LOSP particles for $a=3$(left panel) and $a=2$(right panel) in scenario II.}
\label{fig6}
\end{figure}

The freeze out relic density for various LOSP are shown in figure \ref{fig6}. Again, the lightest neutralino $\tl{\chi}_1^0$ (in $a=3$ case) LOSP can not be the DM candidate because its relic abundance will over close the universe. If the axino is the LSP and act as the DM particle, the LOSP can later decay into axino after its freezing out.

%%%%%%%%%%%%%%%%%%%%%%%%%%Fig1%%%%%%%%%%%%%%%%%%
%%%%%%%%%%%%%%%%%%%%%%%%%%%%%%%%%%%%%%%%%%%%%%%%%%%%%%
%%%%%%%%%%%%%%%%%%%%%%%%%Fig1%%%%%%%%%%%%%%%%%%
%%%%%%%%%%%%%%%%%%%%%%%%%%%%%%%%%%%%%%%%%%%%%%%%%%%%%
    %%%%%%%%%%%%%%%%%%%%%%%%%%Fig1%%%%%%%%%%%%%%%%%%
%%%%%%%%%%%%%%
\section{\label{secV}Conclusions}

We propose a minimal Yukawa deflection scenario of AMSB from the Kahler potential through the Higgs-messenger mixing. Salient features of this scenario are discussed and realistic MSSM spectrum can be obtained. Such a scenario, which are very predictive, can solve the tachyonic slepton problem with less messenger species. Numerical results indicate that the LOSPs predicted by this scenario can not be good DM candidates. So it is desirable to extend this scenario with a Peccei-Quinn sector to solve the strong CP problem and at the same time provide new DM candidates. We propose a way to obtain a light axino mass in SUSY KSVZ axion model with (deflected) anomaly mediation SUSY breaking mechanism. The axino can possibly be the LSP and act as a good DM candidate.

\begin{acknowledgments}
We are very grateful to the referee for goods suggestions. This work was supported by the Natural Science Foundation of China under grant numbers 11675147,11775012.
\end{acknowledgments}

\end{document}